\documentclass[10pt,aps,showpacs,twocolumn,unsortedaddress,nofootinbib,floatfix]{revtex4-1}

\newcommand{\mike}{\color{black}}

\newcommand{\detlef}[1]{\textcolor{black} {#1}}

\newcommand{\sidd}[1]{\textcolor{black} {#1}}

\usepackage[dvipsnames]{xcolor}
\usepackage{textgreek}
\usepackage{graphicx}
\usepackage{dcolumn}
\usepackage{bm}
\usepackage{array}
\usepackage{physics}
\usepackage[version=4]{mhchem}
\usepackage{mathtools,mathrsfs,amsfonts,bigints,dsfont}
\usepackage[caption=false]{subfig}
\usepackage[percent]{overpic}
\usepackage{tikz,pgfplots}
\usepackage{tikz-3dplot}
\usepackage{rotating}
\usepackage{todonotes}
\usepackage[normalem]{ulem}
\usepackage{tabularx,soul}
\usepackage{relsize}

\usepackage[unicode=true, pdfusetitle, bookmarks=true, bookmarksnumbered=false, bookmarksopen=false, breaklinks=true, pdfborder={0 0 0}, backref=false, colorlinks=true, linkcolor=blue, citecolor=blue, urlcolor=blue]{hyperref}

\usetikzlibrary{plotmarks}
\usetikzlibrary{positioning,spy}
\usepgfplotslibrary{fillbetween}
\usetikzlibrary{decorations.markings}
\usepgfplotslibrary{groupplots}
\allowdisplaybreaks
\pgfplotsset{compat=1.10}

\renewcommand{\pi}{\text{\textpi}}
\renewcommand{\Pi}{\text{\textPi}}
\renewcommand{\nu}{\text{\textnu}}
\renewcommand{\zeta}{\text{\textzeta}}
\renewcommand{\eta}{\text{\texteta}}
\renewcommand{\lambda}{\text{\textlambda}}
\DeclareRobustCommand{\chi}{{\mathpalette\irchi\relax}}
\usepackage{scalerel}

\def\up{{\scalerel*{\uparrow}{1}}}

\def\by{{\scalerel*{/}{1}}}

\def\ep{{\stackrel{\uparrow}{}}}

\def\updown{{\scalerel*{\updownarrow}{1}}}

\def\epdown{{\stackrel{\updownarrow}{}}}

\newcommand{\irchi}[2]{\raisebox{\depth}{$#1\text{\textchi}$}}

\newcommand{\bb}[1]{\pmb{#1}}

\newcommand{\myfrac}[3][0pt]{\genfrac{}{}{}{}{\raisebox{#1}{$#2$}}{\raisebox{-#1}{$#3$}}}

\newcolumntype{P}[1]{>{\centering\arraybackslash}p{#1}}

\pgfkeys{/pgf/declare function={arcsinh(\x) = ln(\x + sqrt(\x^2+1));}}

\begin{document}

\title{Exotic Bohmian arrival times of spin-1/2 particles I--An analytical treatment}

\author{Siddhant Das}
\email{Siddhant.Das@physik.uni-muenchen.de}
\author{Markus N\"{o}th}
\email{noeth@math.lmu.de}
\author{Detlef D\"{u}rr}
\email{duerr@mathematik.uni-muenchen.de}
\affiliation{Mathematisches Institut, Ludwig-Maximilians-Universitat M\"{u}nchen, Theresienstr. 39, D-80333 M\"{u}nchen, Germany
}

\date{\today}

\begin{abstract}
It is well known that orthodox quantum mechanics does not make unambiguous predictions for 
the statistics in arrival time (or time-of-flight) experiments. Bohmian mechanics (or de Broglie-Bohm theory) offers a distinct conceptual advantage in this regard, owing to the well defined concepts of point particles and trajectories embedded in this theory. We revisit a recently proposed experiment [S. Das and D. D\"urr, Sci. Rep. (2019)], {\mike the numerical analysis of which revealed} a striking spin dependence in the \sidd{(Bohmian) time-of-arrival distributions of a spin-1/2 particle}. We present here a mathematically tractable variant of the same experiment, where the predicted effects \sidd{can be established} rigorously. We also \sidd{obtain} some new results that can be compared with experiment.
\end{abstract}

\maketitle

\section{Introduction}\label{intro}
 The description of arrival times of a quantum particle at a detector (e.g., a scintillation screen in the double-slit experimental setup) is an unsettled issue \cite{MUGA,MUGA1,Allcock1,Vona,AhBohm}. It is well known that time is not a quantum observable in the canonical sense of a self-adjoint operator, hence there is no clear (or unique) way to address this problem from first principles of orthodox quantum mechanics. In fact, many  theoretical proposals for the arrival time distribution of a particle (claimed to be) based on orthodox quantum mechanics turn out to be \emph{ambiguous}, and at times even paradoxical \cite{Leavens1,Mielnik}, not to mention {\mike only vaguely connected to experiments \cite{CTOA}.} It is also known that the statistics of standard quantum measurements are given by positive operator valued measures (POVMs, also referred to as generalized observables) on the particle's Hilbert space \cite{DGZOperators}. In principle, specifying the POVM associated with a given arrival time experiment requires a full quantum mechanical analysis of the macroscopic system comprised of the apparatus and the particle. Since this is practically impossible, there have been many attempts to guess a universal POVM or a universal class of POVMs from symmetry or other principles of orthodox quantum mechanics \cite{KijowskiPOVM,Werner,AhBohm,AS,Rodi}. {\mike To our knowledge, none of the POVMs suggested} have been experimentally verified in a serious manner.
 
We shall study in this paper the arrival time problem within the framework of Bohmian mechanics, which \sidd{offers} a broader viewpoint \sidd{on quantum phenomena}, not \sidd{limited by} self-adjoint operators or POVMs. \sidd{More importantly, due to the well defined concepts of point particles and trajectories embedded in this theory, it is naturally suited for computing arrival times of a particle.} We focus on certain special wave functions that can be prepared (e.g., ground states of a potential), and for which the Bohmian arrival time distributions show very striking behaviour. Indeed, the distributions we find are so extremely well articulated that their existence almost demands experimental inspection.

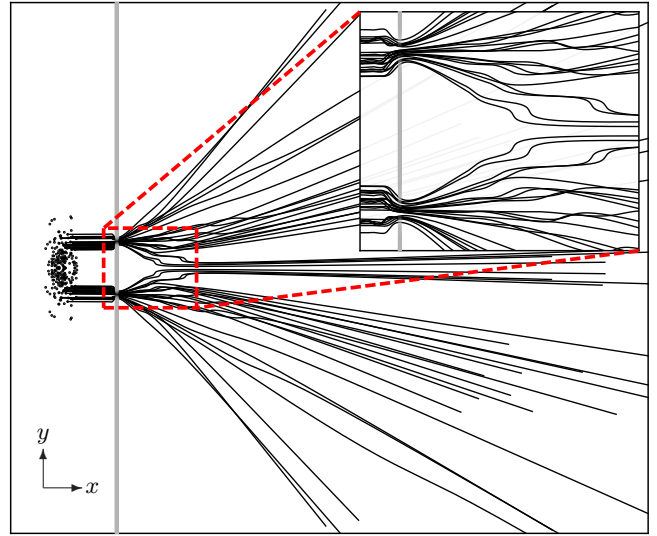
\begin{figure}[!ht]
\centering
\begin{overpic}[width=\columnwidth]{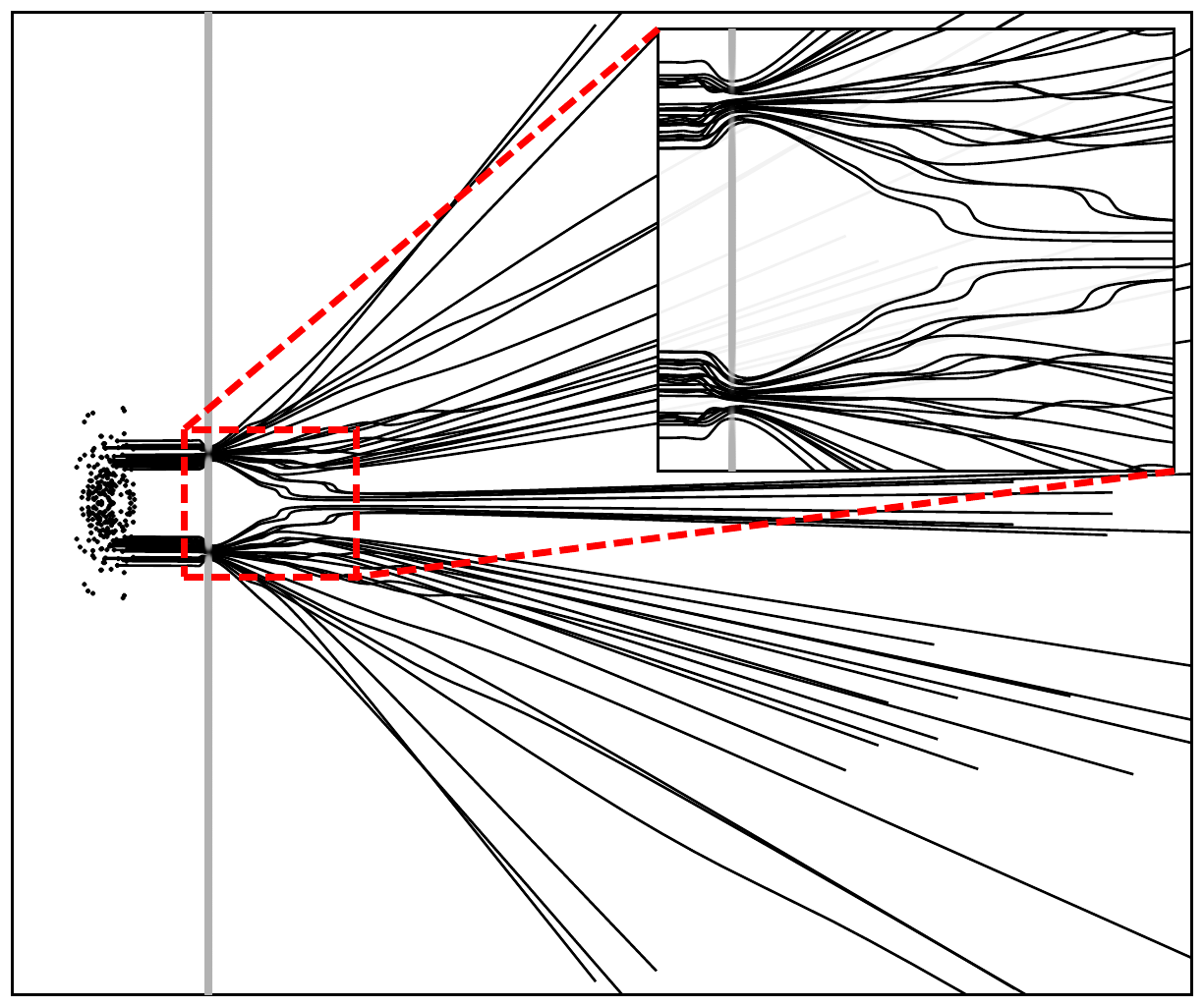}
\put(6,8){\color{Black}\vector(0,1){6}}
\put(6,8){\color{Black}\vector(1,0){6}}
\put (5,15.5) {$y$}
\put (12.5,7.2) {$x$}
\end{overpic}
\caption{A collection of Bohmian trajectories of a spin-0 particle passing through a double-slit interferometer, with initial positions sampled randomly from the initial \(|\psi|^2\!-\)distribution (dots). Most trajectories are reflected back (not shown). Inset: Magnified view of the near field region. Figure courtesy Leopold Kellers \cite{LeoThesis}.}
\label{Leo}
\end{figure}

We refer to the Bohmian arrival time distributions as \emph{ideal} or \emph{intrinsic} distributions, since the influence of the detector is ignored in our theoretical treatment. {\mike Such} an idealization proves to be satisfactory in many applications (e.g., the double-slit experiment, Fig. \ref{Leo}). \detlef{In a follow-up to this paper \cite{complexPot}} we model the influence of a physical detector (via a phenomenological imaginary potential \cite{Allcock2}), which supports our conviction that the ideal arrival time distributions {\mike are in fact} good approximations to the measured ones.


So far, there exist no experimental data for arrival time distributions {\mike other than that obtained} in the ``far field" or scattering regime \cite{Hils,NICE,Kothe,Kaler,Dalibard,Salomon}. In such experiments, the scattered particle after leaving the source travels freely for a \emph{long distance} (compared to the width of its wave function at the time of preparation), and the measured time-of-flight (TOF) of the particle is explained classically, tacitly assuming the validity of Newtonian mechanics. Such treatments are routinely used for fitting TOF data, both in single-particle experiments involving heavy ions (e.g., \ce{^{40}_{20}Ca+}, \ce{^{232}_{90}Th+}) \cite{Kaler,RB} and many-body experiments \cite{Dalibard,Salomon} involving a cloud of \(\approx10^3\) atoms.

The empirical success of semiclassical methods is not altogether surprising from a Bohmian viewpoint, since the emergence of Newtonian behavior in scattering situations is an ubiquitous feature of this theory. In particular, the wave function of a particle in far field (potential-free) regions becomes an approximate plane wave, consequently the Bohmian trajectories become nearly straight lines of \emph{constant} velocity, {\mike similar to the} Newtonian trajectories of a free particle. On the contrary, in the near field (e.g., close to the slits in a double-slit setup, Fig. \ref{Leo}), the particle is influenced by interference of wave packets, causing its trajectory to meander in a non-Newtonian manner. The (Bohmian) arrival time of the particle at a \emph{distant screen} is {\mike thus to a good approximation explained by classical reasoning}, \sidd{ignoring the negligible time spent in the near field region}. Therefore, soliciting deviations from semiclassical methods, theorists (including those approaching the problem from non-Bohmian viewpoints) have recommended ``\emph{moving the detectors closer to the region of coherent wave packet production, or closer to the interaction region}'' \cite[p. 419]{MUGA1}. However, such a relocation may not only disturb the wave function of the particle in an undesirable way \cite{ZenoMuga}, but also require cutting edge time resolution equipment.

Based on these considerations, an arrival time experiment for a spin-1/2 particle was proposed in \cite{DD,SDThesis}, which {\mike had} the distinctive virtue that the particle in the course of its flight {\emph {never}} {\mike moved} freely. {\mike Therefore, the Bohmian arrival time was} not given by a classical formula (as in the far field scattering situations discussed above). Most importantly, in this experiment the non-classical motion was \emph{not} caused by the interference of waves (as in the regions close to the slits), {\mike but was instead} due to the \emph{spin term} found in the  Bohmian guidance law of a spin-1/2 particle (explained below). The {\mike obtained} arrival time distributions revealed a remarkable spin dependence, hitherto unknown. {\mike Furthermore}, all distinguishing features {\mike were} well preserved even with the detector placed at large distances from the source, {\mike hence} the predictions could be checked by present-day experiments. 

\section{A recap of the experiment proposed in \cite{DD}}
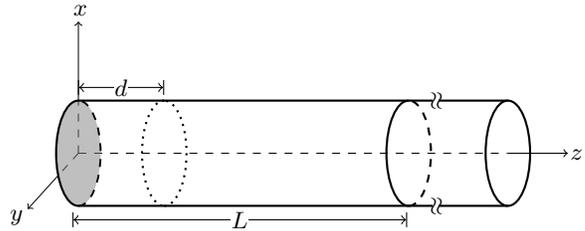
\begin{figure}[!ht]
    \centering
    \def\a{45}
\tdplotsetmaincoords{65}{\a}
\begin{turn}{270}
\begin{tikzpicture}
	[scale=7,
	tdplot_main_coords,
	axis/.style={->,black,very thin},
	curve/.style={black,thin}]
	\def\radius{.1}
	\def\axissize{.25}
	\def\th{0.9}
	\draw[axis] (\radius/1.06418,-\radius/2.9238,0) -- (\axissize/1.06418,-\axissize/2.9238,0) node[anchor=north west,yshift=-3.5mm,xshift=-1mm,rotate=90]{$y$};
	\draw[axis,-,dashed] (0,0,0) -- (\radius/1.06418,-\radius/2.9238,0);
	\draw[axis,-,dashed] (0,0,0) -- (-\radius/1.4142,-\radius/1.4142,0);
	\draw[axis,] (-\radius/1.4142,-\radius/1.4142,0) --(-\axissize/1.4142,-\axissize/1.4142,0) node[anchor=east,yshift=2.3mm,xshift=-1.5mm,rotate=90]{$x$};
	\draw[axis,-,dashed] (0,0,0) -- (0,0,\th);
	\draw[axis] (0,0,\th) -- (0,0,\th+0.5*\axissize) node[anchor=south,rotate=90,xshift=1.2mm,yshift=-2mm]{$z$};
	
	\tdplotsinandcos{\sintheta}{\costheta}{0}
	\tdplotdrawarc[curve,thick]{(0,0,\th)}{\radius*\costheta}{\a-360}{\a}{}{}
	

	\tdplotdrawarc[curve,thick,dotted]{(0,0,\th/5)}{\radius*\costheta}{\a-180}{\a}{}{}
   
    \tdplotdrawarc[curve,thick,dotted]{(0,0,\th/5)}{\radius*\costheta}{\a-360}{\a-180}{}{}
    
	\tdplotdrawarc[curve,thick]{(0,0,\th/1.3)}{\radius*\costheta}{\a-180}{\a}{}{}
    
    \tdplotdrawarc[curve,thick,dashed]{(0,0,\th/1.3)}{\radius*\costheta}{\a-360}{\a-180}{}{}
    
	\tdplotdrawarc[curve,line width=0.25mm,dashed,draw=black, fill=gray, fill opacity=0.5]{(0,0,0)}{\radius*\costheta}{\a}{\a+180}{}{}
	\tdplotdrawarc[curve,thick,draw=black, fill=gray, fill opacity=0.5]{(0,0,0)}{\radius*\costheta}{\a}{\a-180}{}{}
	
	
	\tdplotsinandcos{\sintheta}{\costheta}{\a}
	
	
	\draw[thick] (\radius*\costheta,\radius*\sintheta,0) -- node[fill=white,inner sep=-1.25pt,outer sep=0,anchor=center,yshift=1.9cm]{$\approx$} (\radius*\costheta,\radius*\sintheta,\th);
	
	\draw[|<->|] (1.53*\radius*\costheta,\radius*\sintheta,0) -- node[fill=white,inner sep=0.1pt,outer sep=0,anchor=center,rotate=90] {$L$} (\radius*\costheta,1.53*\radius*\sintheta,\th/1.33);
	
	\draw[|<->|] (-1.5*\radius*\costheta,-\radius*\sintheta,-0.0125) -- node[fill=white,inner sep=0.1pt,outer sep=0pt,anchor=center,rotate=90] {$d$} (-\radius*\costheta,-1.5*\radius*\sintheta,\th/4.7);
	
	\tdplotsinandcos{\sintheta}{\costheta}{\a+180}
	
	\draw[thick] (\radius*\costheta,\radius*\sintheta,0) -- node[fill=white,inner sep=-1.25pt,outer sep=0,anchor=center,yshift=1.9cm]{$\approx$} (\radius*\costheta,\radius*\sintheta,\th);
	
\end{tikzpicture}
\end{turn}
    \caption{Schematic drawing of the experimental setup. The barrier at \(d\) is switched off at \(t=0\) and arrival times are monitored at \(z=L\).}
    \label{waveguide}
\end{figure}

A spin-1/2 particle of mass $m$ is constrained to move in a long waveguide, modeled as a semi-infinite cylinder. Initially, it is trapped between the end face of the waveguide and an impenetrable potential barrier placed at a distance $d$, as shown in Fig. \ref{waveguide}. At the start of the experiment, the particle is prepared in a ground state $\Psi_0$ of this cylindrical box, then the barrier at $d$ is suddenly switched off at, say, $t=0$, allowing the particle to propagate freely within the waveguide. A suitable detector records the {\mike arrival time} (or TOF) $\tau$ of the particle on the \emph{plane} situated a distance $L~(>\!d)$ from the end face of the waveguide. We ask: what is the distribution $\Pi^{\Psi_0}(\tau)$ of these arrival times?

In \cite{DD,SDThesis}, the cylindrical confinement of the waveguide was modelled by a harmonic potential
\begin{equation}\label{vperp}
    V_{\perp}(x,y)=\frac{1}{2}m\,\omega^2(x^2+y^2),
\end{equation}
after popular quadrupole ion traps (a.k.a. Paul traps), while the end face of the waveguide (i.e. the $xy-$plane) and the barrier at $d$ were modelled as hard-wall potential barriers. The ground state wave functions of a spin-1/2 particle confined in such a cylindrical box have the form $\Psi_0(\bb{r})=\psi_0(\bb{r})\chi$, where
\begin{align}\label{space}
    \psi_0(\bb{r})&=\sqrt{\frac{2\kern0.1em m\kern0.1em\omega}{\!\pi\kern0.1em\hbar\kern0.1em d}}\,\theta(z)\,\theta(d-z)\sin(\pi z/d)\,e^{-\frac{m\omega}{2\hbar}(x^2+y^2)}
\end{align}
is the `spatial part' of the wave function, $\chi$ is a normalized two-component spinor ($\chi^{\dagger}\chi=1$), and $\theta(\cdot)$ is Heaviside's step function. 

\par
\begin{figure}
\centering
\begin{overpic}[width=0.95\columnwidth]{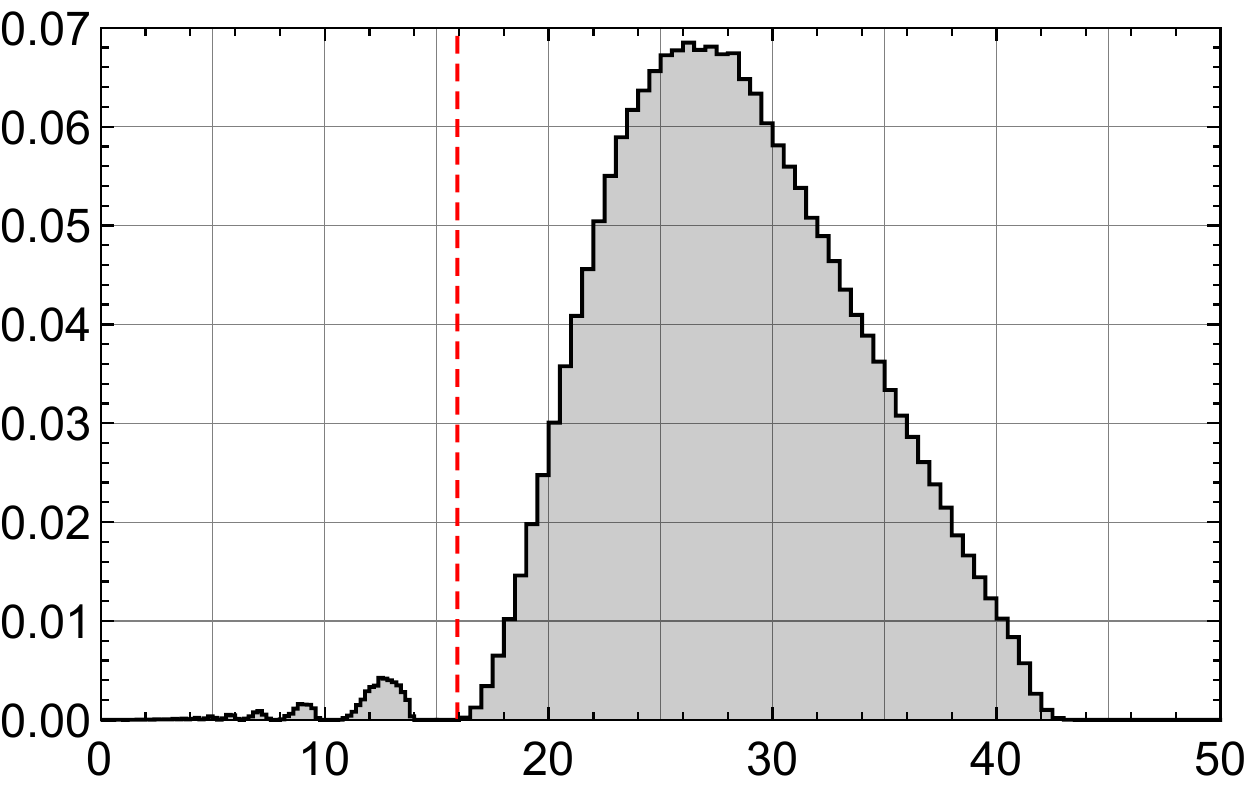}
\put (45.5,-2.75) {$\hbar\,\tau/md^2$}
\put (-5.5,32) {\rotatebox[origin=c]{90}{Relative frequency}}
\put(76.2,9.4){\color{Black}\vector(2,-1){6}}
\put (71,10.7) {$\tau_{\texttt{max}}$}
\put(31,11.2){\color{Black}\vector(1,-1){5}}
\put (28,13) {\small{$\frac{L}{2\pi d}$}}
\end{overpic}
\caption{Arrival time histogram \(\Pi_{\texttt{Bohm}}^{\Psi_0}(\tau) \) {\mike versus (dimensionless) arrival time $\frac{\hbar \tau}{md^2}$}
for the spin up-down wave function: the detector is placed at $L=100\,d$, and we have set \(\beta=0\) with $\omega=10^3\frac{\hbar}{md^2}$. The histogram was generated with $8\times10^5$ Bohmian trajectories whose initial points were randomly drawn from the Born $|\Psi_0|^2-$distribution. All Bohmian trajectories in this case strike the detector before $t=\tau_{\texttt{max}}\approx42.9\,\frac{md^2}{\hbar}$. An infinite train of self-similar smaller lobes, separated by distinct `no-arrival windows', is seen below \(\tau=15.9\,\frac{md^2}{\hbar}\) (dashed line).}
\label{PRLfig}
\end{figure}
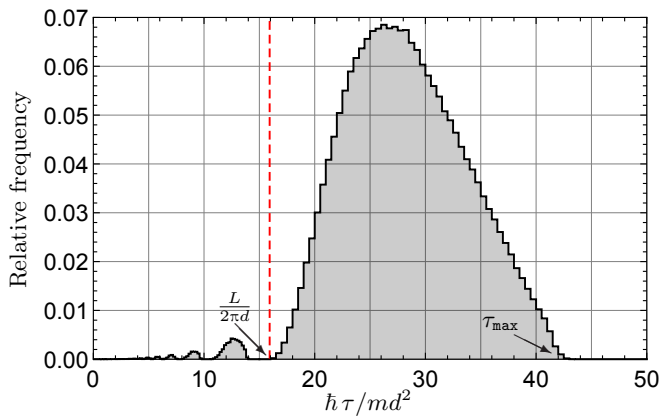

 The instant the barrier is switched off, the wave function spreads dispersively, filling the volume of the waveguide. The particle moves along a definite Bohmian trajectory $\bb{R}(t)=X(t)\,\hat{\bb{x}}+Y(t)\,\hat{\bb{y}}+Z(t)\,\hat{\bb{z}}$ in accordance with Bohm's \emph{guidance law}, eq. \eqref{guidance}, below. For {\mike such an} experimental setup, the \emph{first} arrival time (or hitting time) of a trajectory starting at $\bb{R}_0\equiv\bb{R}(0)$ and arriving at $z=L$ is  
\begin{equation}\label{defn}
    \tau(\bb{R}_0)=\mathrm{min}\!\left\{t\,|\,Z(t,\bb{R}_0)=L,~\bb{R}_0\in\mathrm{supp}(\Psi_0)\right\}\!,
\end{equation}
where $Z(t,\bb{R}_0)\equiv Z(t)$ is the $z-$coordinate of the particle at time $t$, and $\mathrm{supp}(\Psi_0)$ denotes the support of the initial ground state wave function (the region \(0<z<d\)). The arrival time is thus a function of $L$ and the initial position ${\bb{R}_0}$.  The initial positions realized in a sequence of \detlef{ experimental runs} are random, with distribution given by $|\Psi_0|^2$ (see section~\ref{bohmmech}), hence the density of the arrival time distribution
\begin{equation}\label{BohmTime}
    \Pi_{\texttt{Bohm}}^{\Psi_0}(\tau)=\int_{\mathrm{supp}(\Psi_0)}\kern-2.3em\mathrm{d}^3\bb{R}_0~\,\delta\big(\tau(\bb{R}_0)-\tau\big)\,|\Psi_0|^2(\bb{R}_0)\,.\vspace{2pt}
\end{equation}
This distribution {\mike predicted} an unexpected articulated feature for the so-called `up-down' ground state wave function, characterized by
\begin{equation}
    \chi=\frac{1}{\sqrt{2}}\!\left(\!\!\begin{array}{c}1\\e^{i\beta}\\\end{array}\!\!\right)\!,\qquad0\leq\beta<2\pi,
\end{equation}
namely, the density \emph{vanished} beyond a characteristic arrival time $\tau=\tau_{\texttt{max}}$, which we called the `maximum arrival time' (see Fig. \ref{PRLfig}). {\mike That is}, \emph{all} Bohmian trajectories in this case strike the plane $z=L$ before $t=\tau_\texttt{max}$. Furthermore, the distributions for different choices of the parameter $\beta$ \detlef{turned out to be} \emph{identical} (a consequence of the cylindrical symmetry of the waveguide \cite{ddprep}) and {\mike displayed} an infinite sequence of \emph{self-similar} lobes below $\tau=\frac{mdL}{2\pi\hbar}$ (dashed line in Fig. \ref{PRLfig}), which {\mike diminished} in size as $\tau\to0$. It {\mike was also observed} that the smaller lobes are separated by distinct gaps (or `no-arrival windows') inside which the arrival time density is \emph{zero}. Since the predicted distributions {\mike showed} such interesting and significant behavior, we {\mike suggested} that the proposed experiment be performed to test the predictive power of Bohmian mechanics for spin-1/2 particles.

The {\mike observations} in \cite{DD}{\mike, however, were} based on numerical evidence, since analytical solutions for the trajectories were not available. In this paper we explain these features with a mathematically tractable variant of the same experiment. The only modification {\mike is the replacement of} the hard-wall potential barrier at $z=d$ by a smooth harmonic barrier \detlef{\(\frac{1}{2}m\,\omega_\texttt{z}^2z^2\), which \sidd{effectively {\mike limits}} the initial wave function to the region \(0<z<\sqrt{\hbar/m\,\omega_{\texttt{z}}}\),} \sidd{and which is switched off at \(t=0\)}. In this model, the time evolution of the wave function is greatly simplified, and we are able to prove rigorously that the drop-off in the up-down arrival time distribution manifests at a sharply defined time \(\tau_{\texttt{max}}\). \detlef{This supports our conjecture that the notable features reported in \cite{DD} are generic and stable against perturbations.}

We begin in section~\ref{bohmmech} with a brief overview of the Bohmian mechanics of a spin-1/2 particle, applying it in section~\ref{formulation} to {\mike an analysis of the arrival time experiment}. Focusing on two specific wave functions, viz., those of the spin up ($\Psi_\up$) and the spin up-down ($\Psi_\updown$) ground states, we analyze the Bohmian trajectories of the particle following its sudden release from the trap at $t=0$. Arrival time distributions for these cases are found in section~\ref{arrival}, where we also consider their behavior in the limit $\omega\!\to\!\infty$ (i.e. as the diameter of the waveguide goes to zero), \sidd{keeping \(L\) fixed}. The spin up arrival time distribution (being \emph{independent} of $\omega$) remains unaffected, while the up-down arrival time distribution approaches a limiting distribution with $\tau_\texttt{max}\to\omega_\texttt{z}^{-1}\!\sqrt{(m\,\omega_\texttt{z}/\hbar)L^2-1}$. In section~\ref{limiting} we obtain an analytical formula for this limiting distribution, relegating most of the details to the mathematical Appendices. \sidd{On the other hand, both distributions \emph{coincide} in the `no waveguide' limit, \(\omega\to0\).} A confining waveguide is therefore essential to observing this intriguing spin dependence of the arrival time distributions. Section \ref{discussion} offers a general discussion and concludes with a heuristic explanation of the `no-arrival windows' found in Fig. \ref{PRLfig}.

\section{Elements of Bohmian mechanics}\label{bohmmech}
Bohm's theory, like Newtonian mechanics, describes the motion of point particles. However, this dynamics is of first order, therefore the motion of an \emph{isolated} particle is governed by an equation of the type
\begin{equation}\label{guidance}
    \frac{\mathrm{d}}{\mathrm{d}t}\bb{R}(t)=\bb{\mathrm{v}}_{\texttt{Bohm}} (\bb{R}(t),t),
\end{equation}
where $\bb{R}(t)\in\mathbb{R}^3$ is the position of the particle at time $t$, and $\bb{\mathrm{v}}_{\texttt{Bohm}}$ is the velocity field. In other words, Bohmian trajectories are integral curves of this (Bohmian) velocity field, i.e., a solution of eq. \eqref{guidance} for some initial position $\bb{R}_0\equiv\bb{R}(0)$.
\par The Bohmian velocity field for a spin-1/2 particle of mass $m$ is given by \cite[Ch.~10]{HollandPhilippidis,BohmHiley}
\begin{equation}\label{velocity}
    \bb{\mathrm{v}}_{\texttt{Bohm}} (\bb{r},t)=\frac{\hbar}{m}\mathrm{Im}\!\left[\frac{\Psi^{\dagger}\grad\Psi}{\Psi^{\dagger}\Psi}\right]+\frac{\hbar}{2m}\!\left[\frac{\curl\!\left(\Psi^{\dagger}\bb{\sigma}\Psi\right)}{\Psi^{\dagger}\Psi}\right]\!,
\end{equation}
where $\Psi\equiv\Psi(\bb{r},t)$, the wave function, is a two-component complex-valued spinor solution of the Pauli equation 
\begin{align}\label{pauli}
   i\hbar\frac{\partial}{\partial t}&\Psi(\bb{r},t)=-\frac{\hbar^2}{2m}(\bb{\sigma}\!\cdot\!\grad)^2\Psi(\bb{r},t)+V(\bb{r},t)\Psi(\bb{r},t),
\end{align}
with given initial condition $\Psi_0(\bb{r})$, $\Psi^{\dagger}$ is its adjoint, and $\bb{\sigma}=\sigma_\texttt{x}\,\hat{\bb{x}}+\sigma_\texttt{y}\,\hat{\bb{y}}+\sigma_\texttt{z}\,\hat{\bb{z}}$ is a 3-vector of Pauli spin matrices. Here, $V(\bb{r},t)$ denotes an external potential characterizing the interactions of the particle with its surroundings. If magnetic fields are present, the gradient in eq. \eqref{velocity} and \eqref{pauli} should be understood as the gauge covariant derivative, involving the vector potential. 

For almost every $\bb{R}_0$ (w.r.t. the $|\Psi_0|^2$ measure) and under general conditions on the initial wave function $\Psi_0$ and the potential $V$ one has \emph{existence} and \emph{uniqueness} of Bohmian trajectories \cite{Berndl,RodiTufel}. In particular, this implies that Bohmian trajectories \emph{cannot} run into nodes (or zeroes) of the wave function, where the velocity field is ill-defined.  

The dynamical equations (\ref{guidance}-\ref{velocity}) and \eqref{pauli} are both time-reversal\footnote{Unlike a scalar wave function, the time reversal transformation of a spinor is implemented by \(\Psi\rightarrow-i\,\bb{\sigma}_\texttt{y}\Psi^*\) \cite[eq. 4.4.65]{sakurai2010modern}. One can easily verify that under this transformation \(\bb{\mathrm{v}}_{\texttt{Bohm}}\) flips its direction.} and Galilean invariant. Moreover, eq. (\ref{guidance}-\ref{velocity}) is the \emph{unique} nonrelativistic limit of the guidance equation for a relativistic spin-1/2 particle, whose wave function satisfies the Dirac equation \cite{Holland2003,Holland,HollandPhilippidis}.


In Bohmian mechanics the primary role of the wave function is to determine the motion of the particle, while its statistical significance is a derived consequence. As much as in Newtonian mechanics, unique initial conditions lead to unique outcomes via eq. \eqref{guidance} and \eqref{pauli}, nevertheless, experimental predictions made by Bohmian mechanics are always probabilistic in character. This is because the initial particle positions realized in a sequence of identically prepared experiments (i.e. experiments with the same initial wave function $\Psi_0$) are typically random, with distribution given by $|\Psi_0|^2$ (see \cite{DGZBorn} for a justification). By virtue of the velocity field \eqref{velocity}, the position of a Bohmian particle remains $|\Psi|^2$ distributed at \emph{any} later time $t$. This property is known as \emph{equivariance}.

\par {\mike The ostensible randomness, together with the equivariance of the \(|\Psi|^2\) measure and its precise interpretation,} imply that Bohmian predictions {\mike must} agree with the predictions of  orthodox quantum mechanics, \emph{whenever the latter are unambiguous} (e.g. position, momentum, \sidd{and spin} measurements) \cite{BohmHiley,DGZOperators}. As explained in the introduction, there is no clear quantum mechanical prediction for the arrival time statistics of a particle at present. Thus, our analysis provides a possibility to test the predictive power of \sidd{a pragmatic application of} Bohmian mechanics to arrival time experiments.   

\section{Formulation}\label{formulation}
In Cartesian coordinates \(\bb{r}=(x,y,z)\) the cylindrical waveguide depicted in Fig. \ref{waveguide} can be modelled by the potential
\begin{equation}
    V(\bb{r},t)= V_{\perp}(x,y)+V_{\parallel}(z,t),
\end{equation}
where \(V_{\perp}(x,y)=\frac{1}{2}m\,\omega^2(x^2+y^2)\), as in \eqref{vperp}, {\mike but} the axial potential  \(V_{\parallel}(z,t)=v(z)+\theta(-t)\frac{1}{2}m\,\omega_\texttt{z}^2z^2\) is {\mike now} comprised of a harmonic potential barrier (which is switched off at $t=0$) and a hard-wall potential
\begin{equation}\label{wall}
    v(z)=\begin{cases}\infty ~&z\le0  \\ 0 ~&z>0
    \end{cases},
\end{equation}
{\mike delineating} the end face of the waveguide. As the units of measurement of mass, length and time we take, respectively,
\begin{equation*}
    m,\quad\sqrt{\hbar/m\,\omega_\texttt{z}}\kern0.05em,\quad\text{and}\quad1/\omega_\texttt{z}
\end{equation*}
(formally, this amounts to setting $\hbar=m=\omega_\texttt{z}=1$ in all equations). From here on we will work in these units unless otherwise stated. 
\par The ground state wave function of a spin-1/2 particle confined in the trap (for $t<0$) takes the general parametric form
\begin{equation}\label{generic}
    \Psi_0(\bb{r})=\psi_0(\bb{r})\!\left(\!\!\begin{array}{c}\cos(\alpha/2)\\\sin(\alpha/2)\kern0.03em e^{i\beta}\\\end{array}\!\!\right)\!,\quad\begin{array}{c}0\leq\alpha\leq\pi\\[3pt]~\,0\leq\beta<2\pi\\\end{array},
\end{equation}
where
\begin{equation}\label{spatial}
    \psi_0(\bb{r})=A\,\theta(z)\,z\,e^{-\frac{z^2}{2}-\frac{\omega}{2}(x^2+y^2)},
\end{equation}
and $A=\sqrt{4\kern0.1em \omega}/\pi^{3/4}$ is a normalization constant. {\mike For the present discussion of arrival times} we focus on two specific wave functions, viz.,
\begin{equation}\label{init}
    \Psi_\up(\bb{r},0)=\psi_0(\bb{r})\!\left(\!\!\begin{array}{c}1\\0\\\end{array}\!\!\right)\!,\quad\Psi_\updown(\bb{r},0)=\frac{\psi_0(\bb{r})}{\sqrt{2}}\!\left(\!\!\begin{array}{c}1\\1\\\end{array}\!\!\right)\!,
\end{equation}
which \sidd{correspond, respectively, to the choices \(\alpha=\beta=0\), and \(\alpha=\frac{\pi}{2}\), \(\beta=0\), in \eqref{generic}, and} will be referred to as the ‘spin up’, and ‘spin up-down’, wave functions, respectively. Here, up and up-down refers to the orientation of the `spin vector'
\begin{equation}\label{spin vector}
    \bb{\text{s}}\coloneqq\frac{1}{2}\frac{\Psi^{\dagger}\bb{\sigma}\Psi}{|\Psi|^2}=\frac{1}{2}\begin{cases}
       \hat{\bb{z}},~&\Psi=\Psi_\up\\
       \hat{\bb{x}},~&\Psi=\Psi_\updown
\end{cases}
\end{equation}
associated with the wave function \(\Psi\), which is aligned parallel (perpendicular) to the waveguide axis in the case of \(\Psi_\up\) (\(\Psi_\updown\)). Arrival time distributions for general $\alpha$ and $\beta$ are discussed in \cite{DD,SDThesis,ddprep}. 

The solutions of the Pauli equation, eq. \eqref{pauli}, with initial conditions \eqref{init}, are (see Appendix \ref{evolpsi} for details)
\begin{equation}\label{tevolv}
    \Psi_\up(\bb{r},t)=\psi_t(\bb{r})\!\left(\!\!\begin{array}{c}1\\0\\\end{array}\!\!\right)\!,\quad\Psi_\updown(\bb{r},t)=\frac{\psi_t(\bb{r})}{\sqrt{2}}\!\left(\!\!\begin{array}{c}1\\1\\\end{array}\!\!\right)\!,
\end{equation}
where
\begin{equation}\label{spatialt}
    \psi_t(\bb{r})=A\,\theta(z)\,\frac{z}{(1+it)^{3/2}}\,e^{-\frac{z^2}{2(1+it)}-\frac{\omega}{2}(x^2+y^2+2it)}.
\end{equation}
We see that both wave functions propagate dispersively, filling the volume of the waveguide. Their axial widths\footnote{\(\Delta_\texttt{z}(t)\coloneqq\sqrt{\expval{z^2}_{\Psi}-\expval{z}^2_{\Psi}}\,\).} $\Delta_\texttt{z}(t)\approx0.47\sqrt{1+t^2}$ increase with time, while the transverse waveguide potential $V_{\perp}(x,y)$ keeps the wave packets from spreading in the lateral directions. Note that both wave functions vanish at $z=0$, respecting the (\sidd{Dirichlet}) boundary condition at the end face of the waveguide. Note as well that
\begin{equation}
    |\Psi_\up(\bb{r},t)|^2=|\Psi_\updown(\bb{r},t)|^2=\psi_t^*\psi_t^{\textcolor{white}{*}},
\end{equation}
hence the statistical distributions of particle \emph{positions} within the waveguide are identical in both cases at any time $t$. \sidd{However, this does {\mike not imply that the arrival time distributions should be identical, since these depend} exclusively on the underlying dynamics encoded in the guidance law.}

We turn now to the Bohmian trajectories, i.e., the solutions of eq. \eqref{guidance}. The first summand on the right-hand side of the Bohmian velocity field \eqref{velocity}, the so-called \emph{convective} velocity, is the same for both wave functions, viz.,
\begin{equation}\label{conv}
    \mathrm{Im}\!\left[\frac{\Psi^{\dagger}\grad\Psi}{\Psi^{\dagger}\Psi}\right]=\frac{t\,z}{1+t^2}\,\hat{\bb{z}},
\end{equation}
and is directed parallel to the axis of the waveguide. Similarly, the second summand (a.k.a. the \emph{spin} velocity) can be calculated explicitly. Since $\curl\bb{\text{s}}=0$ in both cases (cf. eq. \eqref{spin vector}), the spin velocity can be written as
\begin{align}\label{spin}
    \frac{\curl\!\left(\Psi^{\dagger}\bb{\sigma}\Psi\right)}{2\,\Psi^{\dagger}\Psi}&=\grad(\ln |\psi_t|^2) \times \bb{\text{s}}\nonumber
    \\&=\begin{cases}
       -\omega\kern0.03em \big(y\,\hat{\bb{x}}-x\,\hat{\bb{y}}\big),~&\Psi=\Psi_\up\\[6pt]
       \Big(\frac{1}{z}-\frac{z}{1+t^2}\Big)\,\hat{\bb{y}}+\omega y\,\hat{\bb{z}},~&\Psi=\Psi_\updown
\end{cases}.
\end{align}
The particle position at time $t$ is \(\bb{R}(t)=X(t)\,\hat{\bb{x}}+Y(t)\,\hat{\bb{y}}+Z(t)\,\hat{\bb{z}}.\) Its time derivative \(\dot{\bb{R}}(t)\) features in the guidance equation \eqref{guidance}, {\mike the right-hand side of which can be evaluated using eq. \eqref{conv} and \eqref{spin} to obtain the component equations}
\begin{subequations}\label{EOMup}
\begin{align}
    \dot{X}&=-\omega Y,\label{upx}\\
    \dot{Y}&=\omega X,\label{upy}\\
    \dot{Z}&=\frac{t}{1+t^2}Z\label{upz},
\end{align}
\end{subequations}
for the spin up wave function, and
\begin{subequations}\label{EOMud}
\begin{align}
    \dot{X}&=0,\label{udx}\\
    \dot{Y}&=\frac{1}{Z}-\frac{Z}{1+t^2},\label{udy}\\
    \dot{Z}&=\omega Y+\frac{t}{1+t^2}Z,\label{udz}
\end{align}
\end{subequations}
for the spin up-down wave function.

In view of eq. \eqref{EOMup} and \eqref{EOMud}, the reader might be {\mike concerned that the wave function symmetry in the} \(x-\) and \(y-\)coordinates {\mike has been lost in the guidance equations.} However, this should come as no surprise, since the spin vector \(\bb{\text{s}}\) picks out a preferred direction in each case. We proceed next to the solution of these coupled ODEs with initial condition $\bb{R}_0=X_0\,\hat{\bb{x}}+Y_0\,\hat{\bb{y}}+Z_0\,\hat{\bb{z}}$.

\subsection{Bohmian trajectories for $\Psi_{\ep}$}

The differential equation for the $z-$coordinate, eq. \eqref{upz}, is separable and admits a simple solution:
\begin{equation}\label{zt}
    Z(t)=Z_0\sqrt{1+t^2}.
\end{equation}
Moving now to eq. \eqref{upx} and \eqref{upy}, an easy way of solving these is to introduce a complex coordinate
\begin{equation}
    S(t)\coloneqq X(t)+iY(t),
\end{equation}
the time derivative of which is
\begin{align}\label{eta}
    \dot{S}&=\dot{X}+i\dot{Y}=\omega\big(\!-Y+iX\big)=i\omega S.
\end{align}
Equation \eqref{eta} is readily solved:
\begin{equation}\label{soln}
    S(t)=S_0\,e^{i\omega t},\qquad S_0=X_0+iY_0,
\end{equation}
and the desired solutions can be read off from the real and imaginary parts of \eqref{soln}, viz.,
\begin{subequations}
\begin{align}
    X(t)&=X_0\,\cos(\omega t)-Y_0\,\sin(\omega t),\\
    Y(t)&=Y_0\,\cos(\omega t)+X_0\,\sin(\omega t).
\end{align}
\end{subequations}
From eq. \eqref{soln} we also see that $|S(t)|^2=X^2(t)+Y^2(t)=|S_0|^2$ is a constant of motion. The angular velocity of the particle about the $z-$axis, given by
\begin{equation*}
   \frac{\mathrm{d}}{\mathrm{d}t}\mathrm{Arg}\,[S(t)]=\omega
\end{equation*}
is a constant as well. \sidd{Therefore, a} spin up Bohmian trajectory is a circular \emph{helix} of radius $|S_0|$, which circulates in an anticlockwise sense about the waveguide axis (see Table \ref{summarytab} for an example).

\subsection{Bohmian trajectories for $\Psi_\epdown$}\label{sec:traj_updown}
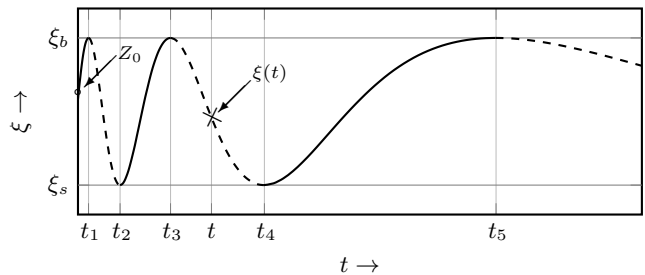
\begin{figure}[!ht] 
\vspace{2mm}
\centering
\setlength{\unitlength}{0.85mm}
\begin{tikzpicture}
\begin{axis}
[
set layers,
width=1.05\columnwidth,
height=0.5\columnwidth,
xlabel = $t\rightarrow$,
xmin = 0.055,
xmax = 0.3,
ylabel = $\xi\rightarrow$,
ymax = 2.9,
ymin = 0.1,
axis line style = thick,
ytick=\empty,
xtick=\empty,
extra y ticks={0.5,2.5},
extra y tick labels={$\xi_s$,$\xi_b$},
extra x ticks={0.05958,0.073297,0.09521,0.113,0.136,0.236640329},
extra x tick labels={$\,t_1$,$\,t_2$,$\,t_3$,$t$,$\,t_4$,$\,t_5$},
xmajorgrids
]

\addplot[domain=0.055:0.3,gray] {2.5};

\addplot[domain=0.055:0.3,gray] {0.5};

\put(0,19.156){\circle{1.03369}}

\put(6.2,24.3){{\scriptsize $Z_0$}}

\put(18.8,14){\rotatebox[origin=c]{-71}{{\large$\times$}}}

\put(27.3,21.4){{\scriptsize $\xi(t)$}}

\addplot[domain=0.05:0.05958,black,thick,smooth] {1.5-sin(deg(1/ln(x+1)))};

\addplot[domain=0.05958:0.073297,black,dashed,thick,smooth] {1.5-sin(deg(1/ln(x+1)))};

\addplot[domain=0.073297:0.09521,black,thick,smooth] {1.5-sin(deg(1/ln(x+1)))};

\addplot[domain=0.09521:0.1357849,black,thick,smooth,dashed] {1.5-sin(deg(1/ln(x+1)))};

\addplot[domain=0.1357849:0.23640329,black,thick,smooth] {1.5-sin(deg(1/ln(x+1)))};

\addplot[domain=0.23640329:0.3,black,thick,smooth,dashed] {1.5-sin(deg(1/ln(x+1)))};

\thinlines

\put(27,21){\vector(-1,-1){5}}

\put(6,24.9){\vector(-1,-1){5}}

\end{axis}
\end{tikzpicture}

\caption{A schematic plot of $\xi(t)$ against $t$, {\mike showing the extreme values} $\xi_s$ and $\xi_b$. Note that $\xi(0)=Z_0$ and \(\dot{\xi}(t)=Y(t)=0\) at the instants \(t_1\), \(t_2,\,\dotsc t_5\). The {\mike solid (dashed) parts of the curve correspond} to $Y(t)>0$ ($<0$).}
\label{xiplot}
\end{figure}

The first equation, eq. \eqref{udx} has the obvious solution
\begin{equation}\label{xcord}
    X(t)=X_0,
\end{equation}
the initial value of \(X\). Consider next eq. \eqref{udy} and \eqref{udz}: these equations are also analytically integrable, however the solutions can only be written in terms of certain nontrivial integrals (i.e. solution by quadrature). Introducing a new function \(\xi\) defined by
\begin{equation}\label{convective}
    Z(t)=\xi(t)\sqrt{1+t^2},
\end{equation}
equations \eqref{udy} and \eqref{udz} can be written as
\begin{subequations}
\begin{align}
    \dot{Y}&=\frac{1}{\sqrt{1+t^2}}\!\left(\frac{1}{\xi}-\xi\right)\label{y}\!,\\
    \dot{\xi}&=\frac{\omega}{\sqrt{1+t^2}}Y\,.\label{xi}
\end{align} 
\end{subequations}
Dividing eq. \eqref{y} by eq. \eqref{xi}, we find
\begin{align}
    \frac{\!\mathrm{d}}{\mathrm{d}t}\Big[\ln\!|\xi(t)|-\frac{1}{2}\,\xi^2(t)-\frac{\omega}{2}Y^2(t)\Big]&=0\nonumber\\[5pt]
    \Rightarrow\ln\xi^2(t)-\xi^2(t)-\omega Y^2(t)&=\text{const.},\label{imp}
\end{align}
{\mike an extremely} useful constant of motion. Since \eqref{imp} holds for all $t$ on the trajectory, one can fix the constant of integration from the initial conditions, i.e.,
\begin{align}
   \ln\xi^2(t)-\xi^2(t)-\omega Y^2(t)&=\ln Z_0^2-Z_0^2-\omega Y_0^2\nonumber\\&\eqqcolon\ln(-g),\label{cons}
\end{align}
noting $\xi(0)=Z_0$ from \eqref{convective}. We have introduced
\begin{equation}
    g=-Z_0^2\,e^{-Z_0^2-\omega Y_0^2}\equiv g(Y_0,Z_0)\label{g}
\end{equation}
for brevity. Solving for $Y$ in \eqref{cons}, we obtain
\begin{align}\label{Y}
    Y(t)= \pm\sqrt{\frac{\ln(\xi^2(t)/\!-g)-\xi^2(t)}{\omega}}\,.
\end{align}
Substitution of \eqref{Y} in \eqref{xi} then yields
\begin{equation}\label{diff}
   \mathrm{sgn}(Y)\,\frac{\mathrm{d}\xi}{\sqrt{\ln(\xi^2/\!-g)-\xi^2}}=\sqrt{\omega}\,\frac{\mathrm{d}t}{\sqrt{1+t^2}},
\end{equation}
where $\mathrm{sgn}(\cdot)$ is the signum function.
\begin{figure}[!ht]
\centering
\begin{tikzpicture}
  \begin{axis}[
      xlabel={$x$}, 
      ylabel={$W$},
      samples=500,
      xmin=-1.2,
      xmax=3.4,
      ymin=-4.2,
      ymax=1.2,
      enlarge y limits=false,
      axis x line=middle,
      axis y line=middle,
      axis line style={-},
      width=0.8\columnwidth,
      height=0.91\columnwidth,
      yticklabel style={xshift=7.5mm},
      ylabel style={xshift=-3mm,yshift=5mm},
      xlabel style={xshift=5mm,yshift=-2.4mm},
      ticklabel style={font=\small}
    ]
    \addplot [red,line width=3pt,opacity=0.4,domain=-0.38:0] (x, 0);
    \addplot [black,dashed,thick,domain=-4.2:-1.01] (x * exp(x), x);
    \addplot [black,thick,domain=-0.95:1.14] (x * exp(x), x);
    \addplot [gray,thin,domain=-4.2:1.2] (-0.38, x);
    \node[font=\fontsize{8}{0}] at (axis cs: -0.8,0.55) {$-1/e$};
    \put(16.3,151){\color{Black}\vector(1,-1){10}}
    \end{axis}
\end{tikzpicture}
\caption{The two real branches of $W(x)$: $W_{-1}(x)$ (dashed); $W_0(x)$ ({\mike solid}). {\mike The thick red line emphasizes the interval $[-1/e, 0)$}, the range of the function $g(Y_0,Z_0)$.}
\label{LambW}
\end{figure}
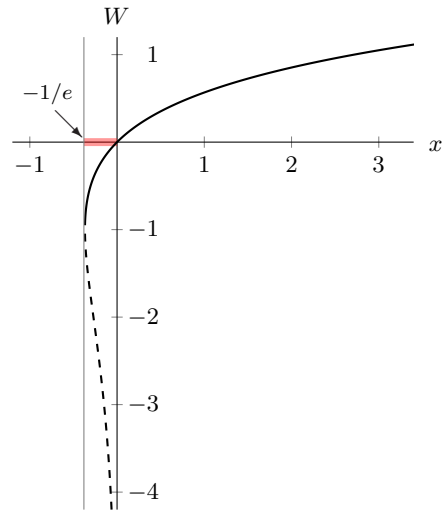
\sidd{In order to integrate the above, we characterize the variation of \(\mathrm{sgn}(Y)\) w.r.t. \(\xi\) as follows:} observe that $\xi(t)$ attains an \emph{extremum} (either a maximum or a minimum) whenever $Y=0$ (cf. eq. \eqref{xi}). These extreme values, {\mike $\bar{\xi}$, therefore} satisfy
\begin{align}
    \ln(\bar{\xi}^{\,2}/\!-g)-\bar{\xi}^{\,2}=0&\Rightarrow-\bar{\xi}^{\,2}\,e^{-\bar{\xi}^{\,2}}=g\nonumber\\&\Rightarrow-\bar{\xi}^{\,2}=W(g)\nonumber\\&\Rightarrow\bar{\xi}=\pm\sqrt{-W(g)},\label{Wg}
\end{align}
where $W(\cdot)$ is the Lambert W function (or product logarithm) \cite{Knuth}. For the values of $g$ permitted by the initial conditions (eq. \eqref{g}), viz.,
\begin{equation}
    -1/e\leq g <0
\end{equation}
there are two possible real values of $W(g)$ (see Fig. \ref{LambW}), denoted by $W_{-1}(g)$ and $W_0(g)$. {\mike Since $\xi(t)>0$},\footnote{The `node evading' property of Bohmian trajectories discussed in Section \ref{bohmmech} implies that the trajectories do not penetrate the base of the waveguide (the \(xy-\)plane), which is a stationary node of \(\Psi_\updown\). Therefore, $Z(t)$ (\,consequently, \(\xi(t)=Z(t)/\sqrt{1+t^2}\,\,\)) \(>0\) for all $t$.} we discard the negative radical in \eqref{Wg}. Thus,
\begin{equation}\label{roots}
    \xi_s\coloneqq\sqrt{-W_0(g)},\qquad\xi_b\coloneqq\sqrt{-W_{-1}(g)},
\end{equation}
satisfying $\xi_s\leq\xi_b$ are the \emph{only} possible extreme values of $\xi(t)$. The following inequality must therefore hold at any given time $t$:
\begin{equation}\label{xiIneq}
    \xi_s\leq\xi(t)\leq\xi_b.
\end{equation}
A schematic plot of $\xi(t)$ is depicted in Fig. \ref{xiplot}. 

\par
\begin{figure}[!ht] 
    \centering
    \begin{overpic}[width=\columnwidth]{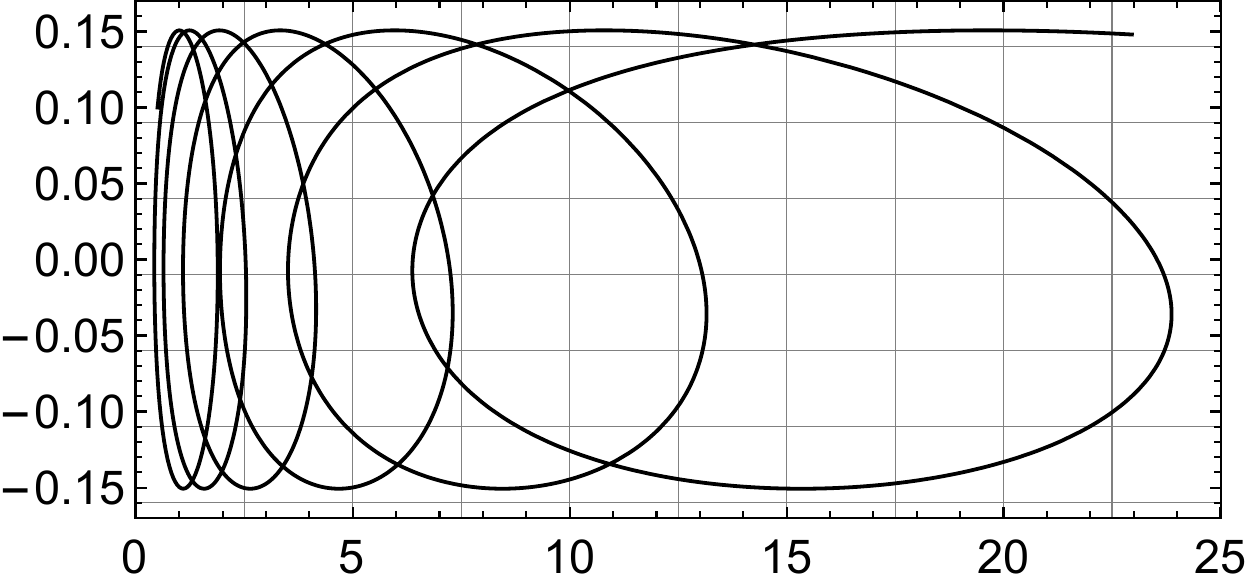}
 \put (53,-3) {\large$z$}
 \put(-2,25.5){\rotatebox[origin=c]{90}{\large$y$}}
\end{overpic}
\caption{A typical Bohmian trajectory of a spin-1/2 particle with wave function $\Psi_\epdown$ and initial position $\bb{R}_0=0.3\,\hat{\bb{x}}+0.1\,\hat{\bb{y}}+0.5\,\hat{\bb{z}}$, \sidd{the \(x-\)coordinate of which is a constant of motion}. The trajectory is plotted for the time interval $[0,20]$ with $\omega=50$.}
\label{udtraj}
\end{figure}

 Since $Y(t)$ changes sign whenever $\xi(t)$ attains an extremum ($Y\!\propto\dot{\xi}$), integrating eq. \eqref{diff} between $t=0$ and some generic time $t$   for the  example {\mike shown} in Fig. \ref{xiplot}, yields 
\begin{align}\label{implicit}
   &\int_{Z_0}^{\xi_b}\frac{\mathrm{d}\xi}{\sqrt{\ln(\xi^2/\!-g)-\xi^2}}-\int_{\xi_b}^{\xi_s}\frac{\mathrm{d}\xi}{\sqrt{\,\cdots}}
 +\int_{\xi_s}^{\xi_b}\frac{\mathrm{d}\xi}{\sqrt{\,\cdots}}\nonumber\\[6pt]
 &\!-\int_{\xi_b}^{\xi(t)}\!\frac{\mathrm{d}\xi}{\sqrt{\,\cdots}}=\int_{Z_0}^{\xi_b}\frac{\mathrm{d}\xi}{\sqrt{\,\cdots}}
 +2\!\int_{\xi_s}^{\xi_b}\frac{\mathrm{d}\xi}{\sqrt{\,\cdots}}
 -\int_{\xi_b}^{\xi(t)}\!\frac{\mathrm{d}\xi}{\sqrt{\,\cdots}}\nonumber\\[6pt]
 &=\sqrt{\omega}\,\sinh^{-1}\!t, 
\end{align}
where we omitted writing the radical explicitly in the integrals above for brevity. Note that a general trajectory attains \(\xi_b\) as the first extremum only if \(Y_0>0\) (as in Fig. \ref{xiplot}), otherwise it attains \(\xi_s\). Consequently, the lower limit of the last integral changes depending on the number of half-cycles \(n\) that are completed between $t=0$ and time $t$. 
If \(n\) is even (e.g., 2, as in \eqref{implicit}), the lower limit of integration of the last term equals the upper limit of integration of the first term. If \(n\) is odd, these limits are different. The generalization of eq. \eqref{implicit} {\mike for any} trajectory may be written as
\begin{widetext}
\begin{subequations}\label{general}
\begin{align}
 &\int_{Z_0}^{\xi_{b}}\frac{\mathrm{d}\xi}{\sqrt{\ln(\xi^2/\!-g)-\xi^2}}
 +n\!\int_{\xi_s}^{\xi_b}\frac{\mathrm{d}\xi}{\sqrt{\,\cdots}}+(-1)^{n+1}\!\!\int_{\frac{\xi_b+\xi_s}{2}+(-1)^n\frac{\xi_b-\xi_s}{2}}^{\xi(t)}~\frac{\mathrm{d}\xi}{\sqrt{\,\cdots}}
 =\sqrt{\omega}\,\sinh^{-1}\!t, &&(Y_0>0)\label{general>0}\\[10pt]
 &\kern-1.3em-\!\int_{Z_0}^{\xi_{s}}\frac{\mathrm{d}\xi}{\sqrt{\ln(\xi^2/\!-g)-\xi^2}}
 +n\!\int_{\xi_s}^{\xi_b}\frac{\mathrm{d}\xi}{\sqrt{\,\cdots}}+(-1)^n\int_{\frac{\xi_b+\xi_s}{2}-(-1)^n\frac{\xi_b-\xi_s}{2}}^{\xi(t)}~\frac{\mathrm{d}\xi}{\sqrt{\,\cdots}}
 =\sqrt{\omega}\,\sinh^{-1}\!t, &&(Y_0<0)\label{general<0}
\end{align}
\end{subequations}
\end{widetext}
where the slightly complicated expressions in the {\mike lower limits of each of the last integrals ensure the correct choice of} \(\xi_b\) or \(\xi_s\) according to the rule explained above. Although  eq. \eqref{general} gives $\xi(t)$ only implicitly, it plays a crucial role in explaining the arrival time statistics of the up-down wave function. Once $\xi(t)$ is found, the complete trajectory of the particle is (implicitly) determined via Eqs. \eqref{convective} and \eqref{Y}. A typical Bohmian trajectory is depicted in Fig. \ref{udtraj}.

\section{Arrival time statistics}\label{arrival}
The first arrival time (or \sidd{passage} time) of a trajectory starting at $\bb{R}_0$ and arriving at $z=L$ is given by eq. \eqref{defn}, where \(\mathrm{supp}(\Psi_0)\), the support of the initial wave function, \sidd{now denotes} the half-space \(z\ge0\).\footnote{Compare this with the ground state \eqref{space} of \cite{DD}, which was supported on the bounded region \(0\le z\le d\).} Since the initial position \(\bb{R}_0\) is $|\Psi_0|^2$-distributed with \(\Psi_0\sim e^{-z^2/2}\)\!, a few initial positions (those with \(Z_0>L\)) will be realized behind the detector surface. Since \(L\gg1\), such spurious initial conditions are rare and can be discarded. We renormalize the arrival time distribution eq. \eqref{BohmTime} accordingly:
\begin{equation}\label{TOFd}
    \Pi_{\up\by\updown}(\tau)=\myfrac[8pt]{\displaystyle\int\limits_{0<Z_0<L}\kern-1em\mathrm{d}^3\bb{R}_0~\delta\big(\tau(\bb{R}_0)-\tau\big)\,|\Psi_{\up\by\updown}(\bb{R}_0,0)|^2}{\displaystyle\int\limits_{0<Z_0<L}\kern-1em\mathrm{d}^3\bb{R}_0~|\Psi_{\up\by\updown}(\bb{R}_0,0)|^2},
\end{equation}
denoting the spin up (spin up-down) arrival time distribution as \(\Pi_{\up}(\tau)\) (\(\Pi_\updown(\tau)\)) for brevity. Here,
\begin{equation}\label{deft}
    \tau(\bb{R}_0)=\mathrm{min}\!\left\{t\,|\,Z(t)=L,~0<Z_0<L\right\}\!.
\end{equation}
 Recalling that
\begin{equation}\label{indist}
    |\Psi_{\up\by\updown}(\bb{R}_0,0)|^2=\frac{4\kern0.1em\omega}{\pi^{3/2}}\theta(Z_0)Z_0^2\,e^{-Z_0^2-\omega(X_0^2+Y_0^2)}
\end{equation}
in both cases (cf. eq. \eqref{init}), the denominator of \eqref{TOFd} can be evaluated explicitly:
\begin{equation}
    \displaystyle\int\limits_{0<Z_0<L}\kern-1em\mathrm{d}^3\bb{R}_0~|\Psi_{\up\by\updown}(\bb{R}_0,0)|^2=\mathrm{erf}(L)-\frac{2L}{\sqrt{\pi}}e^{-L^2}\equiv\lambda_0,
\end{equation}
where \(\mathrm{erf}(\cdot)\) denotes the error function. {\mike In what follows, we} consider the arrival time distributions on a case-by-case basis.

\subsection{Arrival times for \(\Psi_\ep\)}\label{UPArrival}
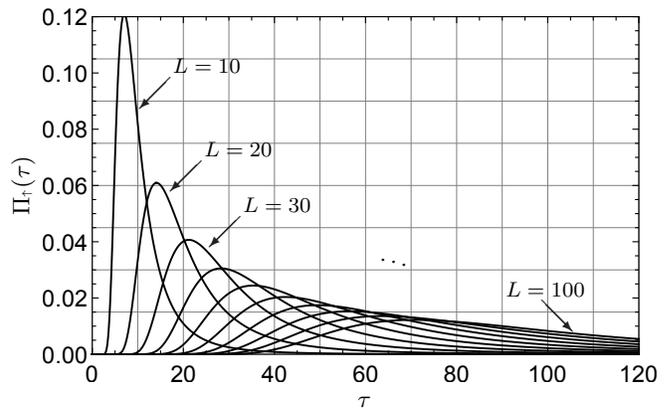
\begin{figure}[!ht]
\centering
\begin{overpic}[width=0.95\columnwidth]{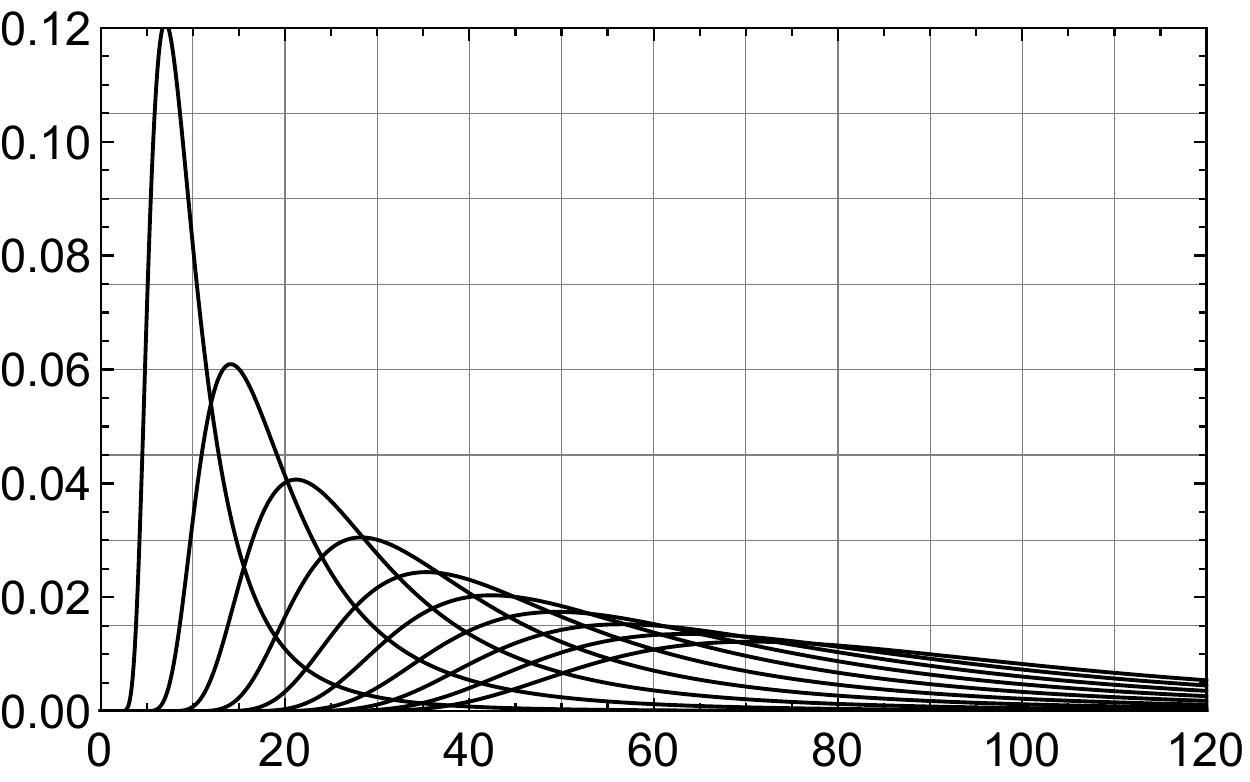}
\put (51,-3) {$\tau$}
\put (-5,32) {\rotatebox[origin=c]{90}{$\Pi_\up(\tau)$}}
\put(20.7,50.4){\color{Black}\vector(-1,-1){5}}
\put (21.1,51.2) {\footnotesize{\(L=10\)}}
\put(25.7,37){\color{Black}\vector(-1,-1){5}}
\put (26.2,37.6) {\footnotesize{\(L=20\)}}
\put(32,28){\color{Black}\vector(-1,-1){5}}
\put (32.5,28.6) {\footnotesize{\(L=30\)}}
\put (55,20) {\rotatebox{-15}{\small{\(\dotsi\)}}}
\put(80.8,14.35){\color{Black}\vector(1,-1){5}}
\put (75,15) {\footnotesize{\(L=100\)}}
\end{overpic}
\caption{Graphs of \(\Pi_{\ep}\) vs. \(\tau\) for select values of \(L\).  The distribution stretches over larger arrival times with increasing \(L\).}
\label{supL}
\end{figure}

The spin up Bohmian trajectories propagate axially outward, {\mike each one crossing} \(z=L\) at most once. Using the exact solution for the trajectory, eq. \eqref{zt} the first arrival time (or simply the arrival time) is given by
\begin{equation}\label{suptau}
    \tau(\bb{R}_0)=\sqrt{(Z_0/L)^2-1}.
\end{equation}
Since this depends only on \(Z_0\), the \(X_0\) and \(Y_0\) integrals in \eqref{TOFd} can be readily evaluated, yielding
\begin{align}\label{remaining}
    \Pi_\up(\tau)&=\frac{4}{\lambda_0\sqrt{\pi}}\!\int_0^L\!\!\!\mathrm{d}Z_0~\delta\!\left(\sqrt{(L/Z_0)^2-1}-\tau\right)\!Z_0^2\,e^{-Z_0^2}.
\end{align}
In order to evaluate the above integral and for later use, we recall \sidd{the identity}
\begin{equation}\label{delta}
    \delta(\phi(\text{x}))=\sum_n\frac{\delta(\text{x}-\text{x}_n)}{|\phi'(\text{x}_n)|},
\end{equation}
where $\text{x}_n$ is a zero of the function $\phi$, $\phi'$ denotes its derivative, and the sum runs over all zeros of $\phi$. For
\begin{equation}
    \phi(Z_0)=\sqrt{(L/Z_0)^2-1}-\tau,
\end{equation}
we obtain two zeros, viz.,
\begin{equation}
    Z_{0\pm}=\pm\frac{L}{\sqrt{1+\tau^2}},
\end{equation}
and evaluating the derivatives of \(\phi\) at \(Z_{0\pm}\), we have
\begin{align}
    \delta\!\left(\sqrt{(L/Z_0)^2-1}-\tau\right)=\frac{\tau Z_0^3}{L^2}\Big[\delta(&Z_0-Z_{0+})\nonumber\\
    &\kern-1em+\delta(Z_0-Z_{0-})\Big].
\end{align}
Since \(Z_{0-}<0\), only the first delta function term \detlef{contributes to the integral and we obtain}:
\begin{equation}\label{updist}
    \Pi_\up(\tau)=\frac{4L^3}{\lambda_0\sqrt{\pi}}\,\frac{\tau\,e^{-\frac{L^2}{1+\tau^2}}}{(1+\tau^2)^{5/2}}.
\end{equation}
Figure \ref{supL} plots \(\Pi_\up(\tau)\) for different values of \(L\).
\sidd{It follows from the above that}
\begin{equation}
    \Pi_\up(\tau)\sim\frac{4L^3}{\lambda_0\sqrt{\pi}}\,\tau^{-4}+\mathcal{O}\big(\tau^{-6}\big),
\end{equation}
as \(\tau\to\infty\). This \sidd{asymptotic behavior} seems to be a characteristic feature of the spin up distribution \cite{DD,ddprep}, and implies that only the mean first arrival time \(\expval{\tau}_\up\), and \(\expval{\tau^2}_\up\) are finite (see Table \ref{summarytab} for exact formulae), while \emph{all} higher moments diverge.

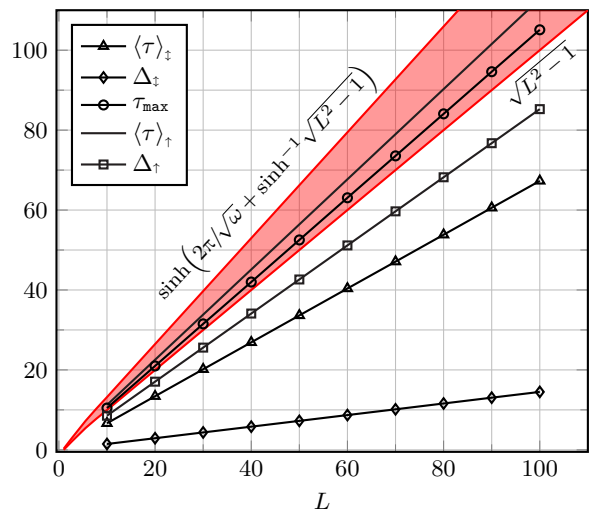
\begin{figure}[!ht]
\centering
\hspace*{-2mm}
\begin{tikzpicture}
\begin{axis}
[
width=\columnwidth,
xlabel = $L$,
xmax = 110,
xmin = -0.5,
ymax = 110,
ymin = -0.5,
ytick = {0,20,40,60,80,100},
xtick = {0,20,40,60,80,100},
minor tick num=1,
axis line style = thick,
legend style = {at={(axis cs:2.5,66)},anchor=south west,thick},
legend cell align = {left},
grid=both
]

\node[font=\fontsize{8}{0}] at (axis cs: 43,66.5) {\rotatebox{47}{$\sinh\!\left(2\pi/\!\sqrt{\omega}+\sinh^{-1}\!\sqrt{L^2-1}\right)$}};

\node[font=\fontsize{8}{0}] at (axis cs: 100,94) {\rotatebox{40}{$\sqrt{L^2-1}$}};

\addplot[name path=A,domain=1:110,red,thick,smooth,forget plot] {sinh(2*pi/sqrt(500)+arcsinh(sqrt(x^2-1)))};

\addplot[name path=B,domain=1:110,red,thick,smooth,forget plot] {sqrt(x^2-1};

\addplot[red,forget plot,opacity=0.4] fill between[of=A and B];

\addplot[
mark = triangle,
thick,
] coordinates {
(10,6.64475)
(20,13.417)
(30,20.1628)
(40,26.8991)
(50,33.6349)
(60,40.3694)
(70,47.0906)
(80,53.8208)
(90,60.566)
(100,67.2964)
};\addlegendentry{$\expval{\tau}_\updown$}

\addplot[
mark = diamond,
thick,
] coordinates {
(10,1.46761)
(20,2.90771)
(30,4.35604)
(40,5.80044)
(50,7.24883)
(60,8.69978)
(70,10.1433)
(80,11.5996)
(90,13.0526)
(100,14.4971)
};\addlegendentry{$\Delta_\updown$}

\addplot[
mark = o,
mark size=1.7pt,
thick,
] coordinates {
(10,10.4612)
(20,20.9873)
(30,31.5167)
(40,41.9907)
(50,52.5236)
(60,63.0683)
(70,73.5513)
(80,84.0691)
(90,94.6119)
(100,105.102)
};\addlegendentry{$\tau_{\texttt{max}}$}

\addplot[
color = Black,
thick,
] coordinates {
(10, 11.2271)
(20, 22.5393)
(30, 33.8326)
(40, 45.1211)
(50, 56.4077)
(60, 67.6933)
(70, 78.9785)
(80, 90.2633)
(90, 101.548)
(100, 112.832)
};\addlegendentry{$\expval{\tau}_\up$}

\addplot[
color = Black,
mark = square,
mark size=1.5pt,
thick,
] coordinates {
(10, 8.54123)
(20, 17.0581)
(30, 25.5804)
(40, 34.1041)
(50, 42.6283)
(60, 51.1528)
(70, 59.6775)
(80, 68.2022)
(90, 76.727)
(100, 85.2518)
};\addlegendentry{$\Delta_\up$}

\end{axis}
\end{tikzpicture}

\caption{Graphs of mean first arrival time \(\expval{\tau}\) and standard deviation \(\Delta\) vs. \(L\) for a fixed \(\omega=500\). The spin up statistics, unlike the up-down ones, are independent of \(\omega\). The maximum arrival time $\tau_{\texttt{max}}$ of the up-down distribution is also depicted here, which lies in the shaded region permitted by inequality \eqref{bound on tau}.}
\label{statL}
\end{figure}

Note as well that \(\Pi_\up(\tau)\) is independent of the trapping frequency \(\omega\), which dropped out obligingly in eq. \eqref{remaining}. \detlef{The reason is that} the motion in the \(z-\)direction decouples from the evolution of the \(x-\) and \(y-\)coordinates of the particle (cf. eq. \eqref{EOMup}). The arrival time of any trajectory thus depends only on \(Z_0\). Furthermore, since the initial wave function \(\Psi_\up\) is separated in the position coordinates, \(Z_0\) is distributed independently with density \(4\kern0.05em Z_0^2\,\theta(Z_0)\exp(-Z_0^2)/\!\sqrt{\pi}\), which is also independent of \(\omega\).

\subsection{Arrival times for \(\Psi_\epdown\)}\label{UPDownArrival}
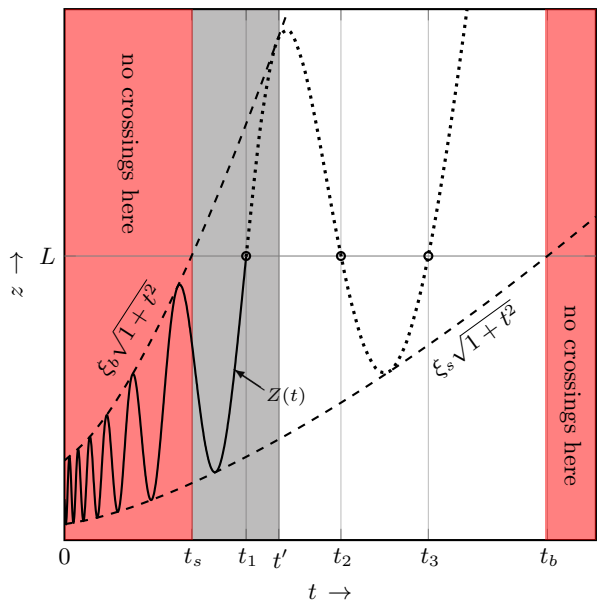
\begin{figure}
\centering
\setlength{\unitlength}{0.8mm}
\begin{tikzpicture}
\begin{axis}
[
set layers,
width=\columnwidth,
height=\columnwidth,
xlabel = $t\,\rightarrow$,
xmin = 0.019,
xmax = 0.21,
ylabel = $z\,\rightarrow$,
ymax = 20,
ymin = 0,
axis line style = thick,
ytick=\empty,
xtick=\empty,
extra y ticks={10.7},
extra y tick labels={$L$},
extra x ticks={0.019,0.06468,0.08428,0.096,0.118354,0.149773,0.19256},
extra x tick labels={$0$,$\,t_s$,$\,t_1$,$\,t^{\prime}$,$\,t_2$,$\,t_3$,$\,t_b$},
xmajorgrids
]

\path [fill=red,opacity=0.5] (0,0) rectangle (46,200);

\path [fill=red,opacity=0.5] (173,0) rectangle (200,200);

\path [fill=Black,opacity=0.3] (46,0) rectangle (77.5,200);

\addplot[domain=0.019:0.21,gray] {10.7};

\put(33.5,23.2){{\scriptsize $Z(t)$}}

\node[label=below:\rotatebox{57}{\(\xi_b\sqrt{1+t^2}\)}] at (23,105) {};

\node[label=below:\rotatebox{-90}{no crossings here}] at (23,190) {};

\node[label=below:\rotatebox{-90}{no crossings here}] at (182,95) {};

\put(60,32){\rotatebox[origin=c]{36.5}{{$\xi_s\sqrt{1+t^2}$}}}

\addplot[domain=0.019:0.08428,black,thick,smooth,samples=600] {(1.5-sin(deg(1/ln(x+1))))*sqrt(1+(40*x)^3)};

\addplot[domain=0.08428:0.21,black,very thick,smooth,dotted] {(1.5-sin(deg(1/ln(x+1))))*sqrt(1+(40*x)^3)};

\addplot[domain=0.019:0.21,black,thick,smooth,dashed] {0.5*sqrt(1+(40*x)^3)};

\addplot[domain=0.019:0.21,black,thick,smooth,dashed] {2.5*sqrt(1+(40*x)^3)};

\addplot[
only marks,
mark=o,
thick,
mark size=1.5pt
] coordinates {
(0.08428, 10.7)
(0.118354, 10.7)
(0.149773, 10.7)
};

\thinlines

\put(33.2,25){\vector(-3,2){5}}


\end{axis}
\end{tikzpicture}

\caption{A schematic plot of \(Z(t)\) vs. \(t\) for a spin up-down Bohmian trajectory, enveloped between the dashed curves \(\xi_s\sqrt{1+t^2}\) and \(\xi_b\sqrt{1+t^2}\). The trajectory intersects \(z=L\) at the instants \(t_1\) (\(=\tau\)), \(t_2\) and \(t_3\), which lie in the interval \([t_s,t_b]\), in accordance with \eqref{basicIneq}.}
\label{Zplot}
\end{figure}

In this case an explicit formula for \(\tau(\bb{R}_0)\), such as \eqref{suptau} cannot be found, as the Bohmian trajectories are known only implicitly (cf. Section \ref{sec:traj_updown}). Moreover, considering the quasiperiodic character of \(\xi\) (Fig. \ref{xiplot}), a typical Bohmian trajectory intersects the plane \(z=L\) multiple times, as shown in Fig. \ref{Zplot}. Experimentally, of course, only the first crossing time \(t_1\) (\(=\tau\)) is relevant {\mike (the time at which the particle is detected)}.
However,  {\mike \(Z(t_k)=L\) at {\emph {any}} crossing time $t_k$, which as a result of eq. \eqref{convective} and inequality \eqref{xiIneq} implies
\begin{equation}
    \xi_s\leq\frac{L}{\sqrt{1+t_k^2}}\leq\xi_b.
\end{equation}
Solving for \(t_k\) above, keeping in mind that
\begin{align}\label{important}
    0<\xi_s\leq1,\quad\text{and}\quad
    1\leq\xi_b<\infty
\end{align}
(cf. eq. \eqref{roots}), yields an analogous inequality for any crossing time of a Bohmian trajectory:
\begin{equation}\label{basicIneq}
    t_s\leq t_k\leq t_b\,,
\end{equation}
where}
\begin{equation}\label{tbounds}
   t_s\coloneqq \theta(L-\xi_b) \sqrt{\frac{L^2}{\xi_b^2}-1}, \qquad t_b\coloneqq\sqrt{\frac{L^2}{\xi_s^2}-1}\,.
\end{equation}
That is, any crossing of a given trajectory, including the first one, commences before time \(t_b\). However, recalling eq. \eqref{g}, one finds that \(g\) approaches zero whenever \(Y_0\) or \(Z_0\) become very large, or even when \(Z_0\approx0\), consequently \(\xi_s=\sqrt{-W_0(g)}\) also approaches zero (see Fig \ref{LambW}). For such initial conditions, \(t_b\) gets arbitrarily large, hence does not explain the uniform upper bound on the arrival times (\(\tau_\texttt{max}\)) found in Fig. \ref{PRLfig}. 

To derive such a bound, consider the \emph{first} instant after \(t=t_s\), say \(t^\prime\), at which a given trajectory touches the upper envelope \(\xi_b\sqrt{1+t^2}\), depicted in Fig. \ref{Zplot}. At this instant, \(Z(t^\prime)=\xi_b\sqrt{1+t^{\prime 2}}\), and since \(t^\prime>t_s\),
\begin{equation*}
    Z(t^\prime)> \xi_b \sqrt{1+t^2_{\text{s}}}\geq L,
\end{equation*}
substituting the definition of \(t_s\) from eq. \eqref{tbounds}. {\mike Thus, the first crossing $\tau$ necessarily occurs {\emph {before}}} \(t=t^\prime\), and we have
\begin{equation}\label{sandwitch}
    t_s\leq\tau\leq t^\prime.
\end{equation}
Since \(t^{\prime}\) lies within \emph{at most} one full cycle after \(t_s\), subtracting equation \eqref{general} evaluated at \(t=t_s\) from that evaluated at \(t=t^{\prime}\), implies
\begin{equation}\label{hastheintegral}
 \sinh^{-1}\kern-0.05em t^\prime\leq\sinh^{-1}\!t_s+\frac{2}{\!\sqrt{\omega}}\int_{\xi_s}^{\xi_b}\!\! \frac{\mathrm{d}\xi}{\sqrt{\ln(\xi^2/\!-g)-\xi^2}}.
\end{equation}
For any initial condition of the trajectory, the above integral \sidd{remains} bounded:
\begin{align*}
\int_{\xi_s}^{\xi_b}\!\! \frac{\mathrm{d}\xi}{\sqrt{\ln(\xi^2/\!-g)-\xi^2}}
&\le\int_{\xi_s}^{\xi_b}\frac{\mathrm{d}\xi}{\sqrt{(\xi-\xi_s)(\xi_b-\xi)}}\\
&=\int_{\xi_s}^{\xi_b}\frac{\mathrm{d}\xi}{\sqrt{\left(\!\frac{\xi_b-\xi_s}{2}\!\right)^{\!\!2}-\left(\!\xi-\frac{\xi_b+\xi_s}{2}\!\right)^{\!\!2}}}\\
&=\!\int_{-1}^1 \frac{\mathrm{d}u}{\sqrt{1-u^2}}=\pi,
\end{align*}
substituting \(\xi=\frac{\xi_b-\xi_s}{2}\,u+\frac{\xi_b+\xi_s}{2}\) in the second line above. The remaining term on the right-hand side of \eqref{hastheintegral} is also bounded, since \(t_s\le\sqrt{L^2-1}\) as a result of eq. \eqref{important} and \eqref{tbounds}, thus yielding
\begin{equation}\label{bound on tau}
\tau \le \sinh \!\left(\frac{2\pi}{\sqrt{\omega}} + \sinh^{-1}\!\sqrt{L^2-1} \right)\!,
\end{equation}
via \eqref{sandwitch}. The first crossing time of \emph{any} Bohmian trajectory is therefore bounded from above. Hence, irrespective of the initial position, the particle strikes the plane \(z=L\) before a maximum arrival time \(\tau_\texttt{max}\). 

\begin{figure}[!ht] 
\centering
\begin{overpic}[width=0.95\columnwidth]{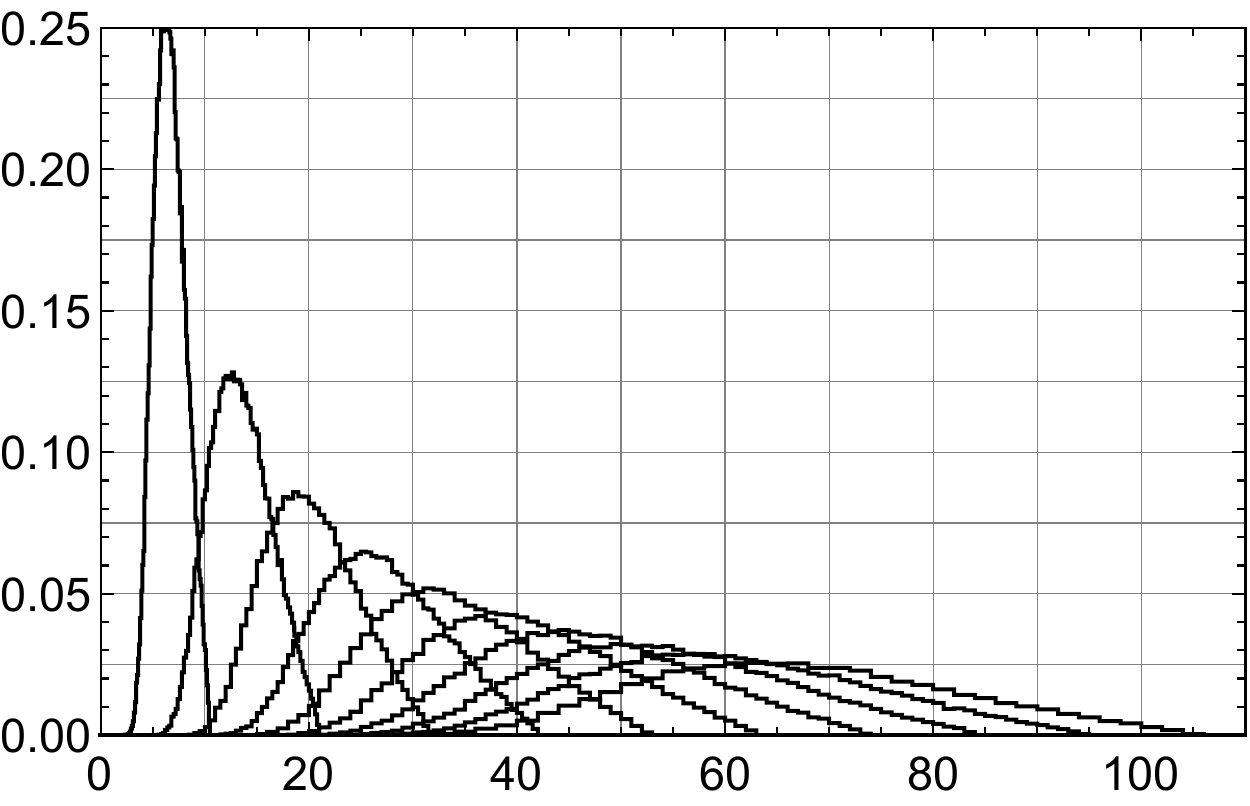}
\put (51,-3) {$\tau$}
\put (-5,33.3) {\rotatebox[origin=c]{90}{$\Pi_\updown(\tau)$}}
\put(20.7,50.4){\color{Black}\vector(-1,-1){5}}
\put (21.1,51.2) {\footnotesize{\(L=10\)}}
\put(25.7,37){\color{Black}\vector(-1,-1){5}}
\put (26.2,37.6) {\footnotesize{\(L=20\)}}
\put(32,28){\color{Black}\vector(-1,-1){5}}
\put (32.5,28.6) {\footnotesize{\(L=30\)}}
\put (55,20) {\rotatebox{-15}{\small{\(\dotsi\)}}}
\put(81,14.6){\color{Black}\vector(-1,-1){5}}
\put (81.5,15.2) {\footnotesize{\(L=100\)}}
\end{overpic}
\caption{Up-down arrival time histograms for select values of \(L\) and \(\omega=500\). Each histogram is {\mike constructed} from \(\approx10^5\) Bohmian trajectories, whose initial conditions were sampled randomly from the initial \(|\Psi|^2-\)distribution \eqref{indist}. It should be noted that for every \(L\) there exists a maximal arrival time \(\tau_\texttt{max}\).}
\label{pii}
\end{figure}
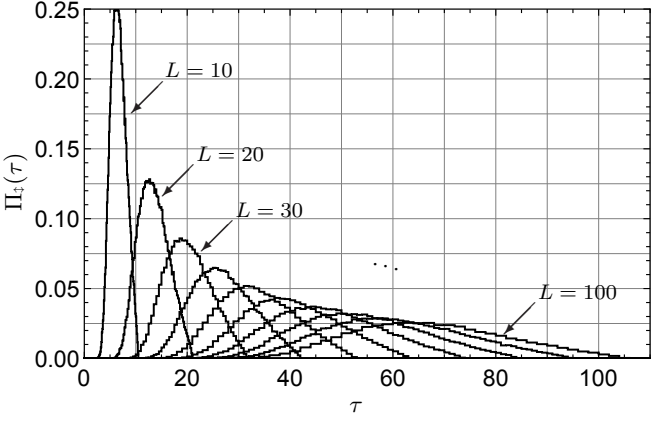

To illustrate this better, we sample \(N\approx10^5\) random initial positions from the \(|\Psi_0|^2-\)distribution \eqref{indist}, solve the up-down equations of motion eq. \eqref{EOMud} numerically for each point in this ensemble, continuing until the trajectory hits \(z=L\), then record the arrival time and plot the histogram for \(\Pi_\updown(\tau)\), Fig. \ref{pii}. Note that a \(\tau_\texttt{max}\) occurs regardless of \(L\). Figure \ref{statL} plots the mean \(\expval{\tau}_\updown\), standard deviation \(\Delta_\updown\), and \(\tau_\texttt{max}\) of these histograms against \(L\). Indeed, \(\tau_\texttt{max}\) lies well below the threshold permitted by \eqref{bound on tau}. 

\subsection{Trapping frequency limits}\label{limiting}
The trapping frequency \(\omega\) measures the effective diameter of the waveguide, which we take to be the width of the radial wave function, viz., \(\sqrt{\hbar/m\,\omega}\) (\(=1/\!\sqrt{\omega}\) in our dimensionless units); typical particle trajectories also lie within this distance from the waveguide axis. {\mike We consider here} the behavior of the arrival time distributions with changing \(\omega\), for a fixed \(L\). As noted {\mike at the} end of Section \ref{UPArrival}, the spin up distribution {\mike is independent of} \(\omega\), so in what follows we focus on the spin up-down {\mike distribution}.

In the limit \(\omega\to0\), the radial confinement of the waveguide is absent, the distribution \(\Pi_\updown(\tau)\) reduces to the spin up distribution \(\Pi_\up(\tau)\) (eq. \eqref{updist}), while the maximum arrival time \(\tau_\texttt{max}\) is pushed to infinity. This can be seen from eq. \eqref{udz}, which for small \(\omega\) approaches its spin up analogue, eq. \eqref{upz}. The latter led directly to the spin up distribution in Section \ref{UPArrival}. However, in this limit, the respective Bohmian trajectories remain manifestly different: The spin up trajectories are straight lines running parallel to the \(z-\)axis, while the spin up-down trajectories take the form \(X(t)=X_0\), \(Y(t)\approx Y_0+(1/Z_0-Z_0)\sinh^{-1}\!t\), \(Z(t)\approx Z_0\sqrt{1+t^2}\).

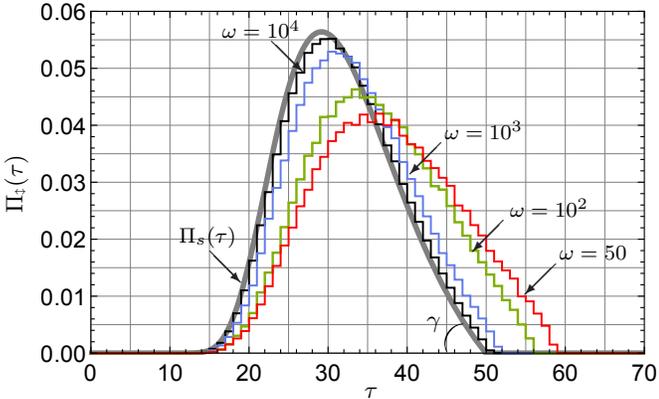
\begin{figure}[!ht]
\centering
\hspace*{-2mm}
\begin{overpic}[width=0.95\columnwidth]{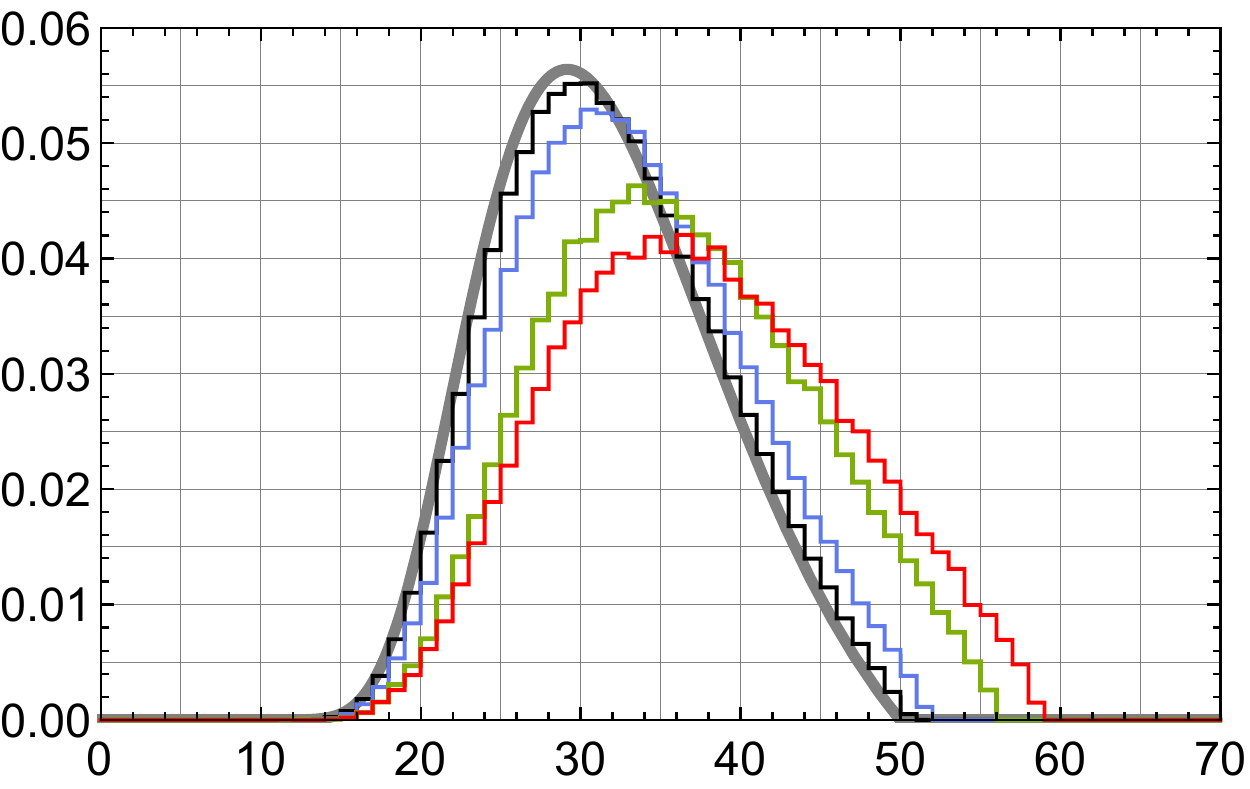}\label{convergence a}
\put (51,-1.7) {$\tau$}
\put (-7.2,32) {\rotatebox[origin=c]{90}{$\Pi_\updown(\tau)$}}
\put(36,56){\color{Black}\vector(1,-1){5}}
\put(28,56.5){\footnotesize{\(\omega=10^4\)}}
\put(63,4.9){\rotatebox{60.5}{\Large\(\frown\)}}
\put(61.1,9.7){\(\gamma\)}
\put(73.5,27){\color{Black}\vector(-1,-1){5}}
\put (74.6,27.6) {\footnotesize{\(\omega=10^2\)}}
\put(82,20){\color{Black}\vector(-1,-1){5}}
\put (82.6,20.6) {\footnotesize{\(\omega=50\)}}
\put(63.4,39.55){\color{Black}\vector(-1,-1){5}}
\put (64,40.14) {\footnotesize{\(\omega=10^3\)}}
\put(26,22){\color{Black}\vector(1,-1){5}}
\put(21,23.4){\footnotesize{\(\Pi_s(\tau)\)}}
\end{overpic}
\caption{ Up-down arrival time histograms for select values of \(\omega\) and $L=50$. The histograms approach \(\Pi_s(\tau)\), the distribution of \(t_s\) (thick gray curve), as \(\omega\to\infty\), while \(\tau_\texttt{max}\to\sqrt{L^2-1}\approx50\). The angle subtended at the foot of the distribution, the podal angle, \(\gamma\approx\tan^{-1}(4.16/L^2)\).}
\label{convergence}
\end{figure}

On the other hand, in the limit \(\omega\to\infty\), the wave function gets compressed onto the waveguide axis, effectively fusing the trajectories onto the same. Even in this rather singular limit, the up-down arrival time distribution converges to a well-defined distribution, a feature illustrated numerically in Fig. \ref{convergence}. 
This behavior can be anticipated from the combined inequality (\ref{sandwitch}-\ref{hastheintegral}):
\begin{equation}\label{bounds}
    t_s\leq\tau\leq\sinh\!\left(\!\frac{2\pi}{\sqrt{\omega}}+\sinh^{-1}\!t_s\right)\!,
\end{equation}
which {\mike suggests} that the first arrival time \(\tau\) approaches \(t_s\), as \(\omega\to\infty\). However, this has to be taken cum grano salis, since \(t_s\) itself depends intricately on \(\omega\) and \(\bb{R}_0\). Therefore, we consider the convergence of \(\tau\to t_s\) \emph{in distribution}.

From the left inequality of \eqref{bounds}, we have
\begin{align}\label{pp}
    P(\tau\le t)=P(t_s\le \tau~\text{and}~t_s\le t)\le P(t_s\le t),
\end{align}
\(P(\cdot)\) is the Born probability, given by the \(|\Psi_0|^2\) measure. Now, using the right inequality of \eqref{bounds}, rewritten as
\begin{equation}\label{name}
    \sinh\!\left(\sinh^{-1}\!\tau-\frac{2\pi}{\sqrt{\omega}}\!\right)\le t_s,
\end{equation}
we have for a given \(t\),
\begin{align}
    &P\!\left(t_s\le\sinh\!\left(\sinh^{-1}\!t-\frac{2\pi}{\sqrt{\omega}}\!\right)\right)\nonumber\\[6pt]
    &=P\!\left(\eqref{name}~\,\text{and}\,~t_s\le \sinh\!\left(\sinh^{-1}\!t-\frac{2\pi}{\sqrt{\omega}}\!\right)\right)\nonumber\\[7pt]
    &\le P\!\left(\sinh\!\left(\sinh^{-1}\!\tau-\frac{2\pi}{\sqrt{\omega}}\!\right)\le \sinh\!\left(\sinh^{-1}\!t-\frac{2\pi}{\sqrt{\omega}}\!\right)\right)\nonumber\\[7pt]
    &= P(\tau\le t).
\end{align}
Combining the above with \eqref{pp}, yields
\begin{equation}\label{Markus}
    P\!\left(t_s\le\sinh\!\left(\sinh^{-1}\!t-\frac{2\pi}{\sqrt{\omega}}\!\right)\right)\le P(\tau\le t)\le P(t_s\le t).
\end{equation}
To take the limit \(\omega\to\infty\) in \eqref{Markus}, we observe that
\begin{equation}\label{CDF}
    P(t_s\leq t)=\int_0^t\!\!\mathrm{d}t^\prime\,\Pi_s(t^\prime),
\end{equation}
where \(\Pi_s\) is the density of \(t_s\). As shown in Appendix \ref{cherish}, \(\Pi_s\) is \emph{independent} of \(\omega\), thus \(P(t_s\leq t)\) is unaffected in the limit \(\omega\to\infty\). As a result,
\begin{equation}
    \lim_{\omega\to\infty}P(\tau\le t)=\int_0^t\!\!\mathrm{d}t^\prime\,\Pi_s(t^\prime),
\end{equation}
and formally, \(\Pi_\updown(t)\,(\coloneqq\mathrm{d}/\mathrm{d}t P(\tau\leq t))\to \Pi_s(t)\), explaining Fig. \ref{convergence}.

\begin{figure}[!ht]
\centering
\setlength{\unitlength}{0.85mm}
\begin{tikzpicture}
\begin{axis}
[
set layers,
width=\columnwidth,
height=0.69\columnwidth,
xlabel = $L$,
ymode=log,
xmax = 105,
xmin = 0,
ylabel = $\gamma$ (\texttt{rad}),
ymax = 0.5,
ymin = 0,
minor x tick num=2,
axis line style = thick,
grid=both
]

\addplot[only marks, thick, mark = *, mark size=1.5pt, mark options={fill=White}] coordinates {
(10, 0.057637962537026036)
(20, 0.01492658028882372)
(30, 0.006669195771498648)
(40, 0.00364920833041564)
(50, 0.002321821688011882)
(60, 0.0015678134931020476)
(70, 0.0011325625026598395)
(80, 0.0009321658850491813)
(90, 0.0007144330042902254)
(100, 0.0006099650902230471)
};\addlegendentry{$~\omega=500$}

\addplot[only marks, thick, mark = triangle, mark size=1.5pt, mark options={fill=White}] coordinates {
(10, 0.04968548359834203)
(20, 0.012535622016734124)
(30, 0.005563321452564597)
(40, 0.003027867913858335)
(50, 0.001970482114404193)
(60, 0.00136221260665079)
(70, 0.0010260928477861353)
(80, 0.0007873986464391091)
(90, 0.0006517878977285906)
(100, 0.0004913646271161304)
};\addlegendentry{$~\omega=10^4$}
\addplot[domain=1:105,black,thick,smooth] {atan(4.16/x^2)*(pi/180)};
\end{axis}
\end{tikzpicture}
\caption{A comparison of the theoretically calculated podal angle of the limiting distribution (\(\omega\to\infty\)) with numerical estimates for two (large) values of \(\omega\).}
\label{Podangle}
\end{figure}
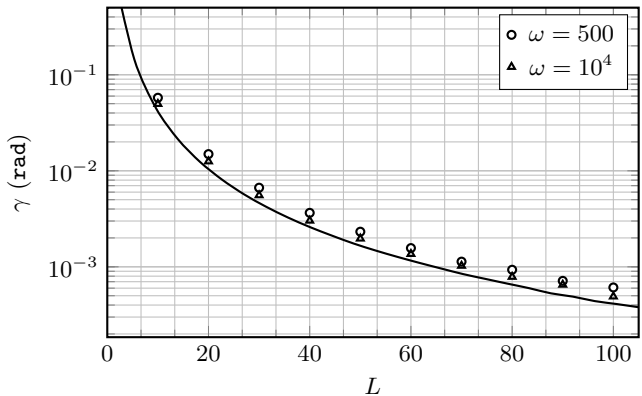

To put this result in perspective, consider a \ce{^{40}_{20}Ca+} ion of mass \(m\approx6.6\times10^{-26}\,\texttt{kg}\), initially trapped in the region \(0\!<\!z\!<\!\sqrt{\hbar/m\,\omega_{\texttt{z}}}\approx10^{-6}\,\texttt{m}\), or \(\omega_\texttt{z}\approx10^4\,\texttt{rad/s}\), and moving in a quadrupole ion trap waveguide. The typical trapping frequencies \(\omega\approx10^7-10^{11}\,\texttt{rad/s}\), which in our dimensionless units correspond to \(\omega\approx10^3-10^7\). For these specifications we will, for all practical purposes, end up with the limiting distribution \(\Pi_s\), as shown in Fig. \ref{convergence}. An explicit formula for this distribution would therefore be very useful. We perform such a calculation in Appendix \ref{cherish}, finding
\begin{align}\label{limdist}
    \Pi_s(\tau)=\frac{\tau L}{(1+\tau^2)^{3/2}}\theta\big(\sqrt{L^2-1}-\tau\big)\Lambda\!\left(\!\frac{L}{\sqrt{1+\tau^2}}\!\right)+\eta\kern0.07em\delta(\tau),
\end{align}
where
\begin{equation}
    \Lambda(x)=\frac{8\kern0.07em x}{\pi\kern0.07em\lambda_0}\kern0.02em(x^2-1)\kern0.05em e^{-x^2}\!\!\!\bigintssss_{\ell(x)}^x\frac{\mathrm{d}u}{\sqrt{2\ln(u/x)+x^2-u^2}},
\end{equation}
\begin{equation}
    \eta\coloneqq\int_L^{\infty}\!\!\!\mathrm{d}x~\Lambda(x),~\text{ and }~ \ell(x)\coloneqq\sqrt{-W_0\big(\!-x^2\kern0.03em e^{-x^2}\big)}\,.
\end{equation}
Note that this distribution vanishes for any \(\tau\!\ge\!\sqrt{L^2-1}\), the limiting value of \(\tau_\texttt{max}\). A tangent line to the distribution at this point defines an angle \(\gamma\) with the \(\tau-\)axis (indicated in Fig. \ref{convergence}) given by
\begin{equation}\label{anggl}
    \gamma\approx\tan^{-1}\!\left(\frac{4.16}{L^2}\right)\!,\,\quad\qquad L\gg1.
\end{equation}
This \emph{podal angle} is a notable characteristic of the up-down distribution. In Fig. \ref{Podangle} we plot numerical estimates of \(\gamma\) against \(L\) for two large values of \(\omega\), obtaining good agreement with equation \eqref{anggl}.

A further surprising feature of the limiting distribution \eqref{limdist} is the appearance of a singular term,  \(\eta\kern0.1em\delta(\tau)\), which implies that a few arrivals occur \emph{instantaneously} in the limit \(\omega\to\infty\). In practice, we cannot observe such arrivals by simply choosing a large value of \(\omega\), and more to the point, initial conditions associated with them are located very near the end face \(z=0\) of the waveguide, hence are atypical. 

\section{Concluding remarks}\label{discussion}
Our findings for the spin up and the spin up-down wave functions with all essential details are collected  in Table \ref{summarytab}.

In comparing with results found in \cite{DD} we would like to emphasize the following:
the maximum arrival time \(\tau_\texttt{max}\) reported in \cite{DD} also manifests in the model considered in this paper, and is shown here to be a consequence of certain special dynamical properties of the Bohmian trajectories, namely,
\begin{enumerate}
    \item the natural convection of the trajectories driven by the dispersing wave packet, 
    \item a quasiperiodic oscillation of the variable \(\xi(t)=\frac{Z(t)}{\sqrt{1+t^2}}\), and
    \item a uniform lower bound (over all initial positions) for the \detlef{maximum \(\xi_b\) (cf. \eqref{roots}) of these  oscillations.}
\end{enumerate} 
The confining waveguide certainly plays a key role here, since the oscillations of \(\xi\) are suppressed in the `no waveguide' limit, \(\omega\to0\), and the up-down arrival times approach the spin up ones. The latter satisfy only property~(i). 

For the model considered in \cite{DD} (cf. Fig. \ref{waveguide}), these properties are difficult to verify, as the wave function separates into an infinite collection of tiny ripples near \(z\approx d\) as soon as the barrier is switched off at \(t=0\) \cite{SDThesis,ddprep}. The ripples, in the course of time, develop into wave packets (separated by nodes), each propagating dispersively along the waveguide.\footnote{This remarkable wave phenomenon, know as \emph{diffraction in time} \cite{Moshinsky}, manifests in response to a sudden change in the boundary conditions of the wave function at a given surface (in this case, the plane \(z=d\)). For the harmonic barrier \(\frac{1}{2}m\,\omega_\texttt{z}^2z^2\) considered in this paper, such conditions are not met, consequently no ripples are observed.} Each of the smaller lobes of the arrival time histogram, Fig. \ref{PRLfig}, is caused by the arrival of particles propagating within the support of {\emph {one}} such wave packet. In particular, due to the nodes separating these wave packets, the particle remains within the support of the {\emph {particular}} wave packet for which its random initial position was realized at \(t=0^+\). The nodes move forward in time, carrying the particle along, hence the arrival times are recorded in bunches. 

The Bohmian dynamics within a given wave packet is very similar to the Bohmian dynamics of the waveguide-confined particle studied here, in the sense that the rear node of a given wave packet resembles the waveguide hard wall at \(z=0\), while the frontal node is analogous to the vanishing tail of the wave function \eqref{tevolv}. Continuing the analogy, the results of this paper would suggest the appearance of a `maximum arrival time' for each wave packet. Such a \(\tau_\texttt{max}\) would necessarily be smaller than the time at which the rear node of the preceding wave packet crossed $L$. This is consistent with the formation of  `no arrival windows' found numerically in \cite{DD}, illustrated in Fig. \ref{PRLfig}.

\begin{table*}[]
    \centering
    \begin{tabularx}{\textwidth}{|>{\centering\arraybackslash}m{4cm}|>{\centering\arraybackslash}m{7cm}|>{\centering\arraybackslash}m{\textwidth-4cm-7cm-0.478cm}|}
\hline
\begin{center}Wave function:\end{center} & \begin{center}\(\Psi_\up (\bb{r},t)= \psi_t(\bb{r})\!\begin{pmatrix}1 \\ 0\end{pmatrix} \)\end{center} & \begin{center}\(\Psi_\updown (\bb{r},t)=\frac{\psi_t(\bb{r})}{\sqrt{2}}\!\begin{pmatrix}1 \\ 1\end{pmatrix}\)\end{center} \\
\hline
& & \\ 
Position probability density: \(\,\Psi^{\dagger}\Psi\)
& \(|\psi_t(\bb{r})|^2\) & \(|\psi_t(\bb{r})|^2\) \\
& & \\
\hline
& & \\ 
Spin vector: $\,\bb{\text{s}}=\frac{1}{2}\frac{\Psi^{\dagger}\bb{\sigma}\Psi}{\Psi^{\dagger}\Psi}$&~\(\frac{1}{2}\,\hat{\bb{z}}\) (along waveguide axis) ~&~ \(\frac{1}{2}\,\hat{\bb{x}}\) (perpendicular to waveguide axis)~\\
&& \\
\hline
Guiding equations&
\begin{center}
\(\begin{aligned}
    \dot{X} &=-\omega Y\\
    \dot{Y} &=\omega X\\
    \dot{Z} &=\frac{t}{1+t^2}Z
\end{aligned}\)\end{center}
&
\begin{center}\(\begin{aligned}
    \dot{X}& =0\\
    \dot{Y}& =\frac{1}{Z}-\frac{Z}{1+t^2}\\
    \dot{Z}& =\omega Y+\frac{t}{1+t^2}Z
\end{aligned}\) \end{center}\vspace{0.3cm}\\
\hline
    &   & \\
Constants of motion& \(X^2+Y^2,~\) and \(~X\dot{Y}-Y\dot{X}\) & \(\displaystyle\ln\!\left(\frac{Z^2}{1+t^2}\right)-\frac{Z^2}{1+t^2}-\omega Y^2,~\) and \(~X\)\\
    &   & \\
\hline 
Typical Bohmian trajectories for respective wave functions: \(\bb{R}(0)=0.05\,\hat{\bb{x}}+0.1\,\hat{\bb{y}}+0.2\,\hat{\bb{z}}\) and \(\omega=20\). Both trajectories are plotted for the time interval \([0,2]\). & \begin{center}
\scalebox{0.6}{
\begin{overpic}[width=\columnwidth]{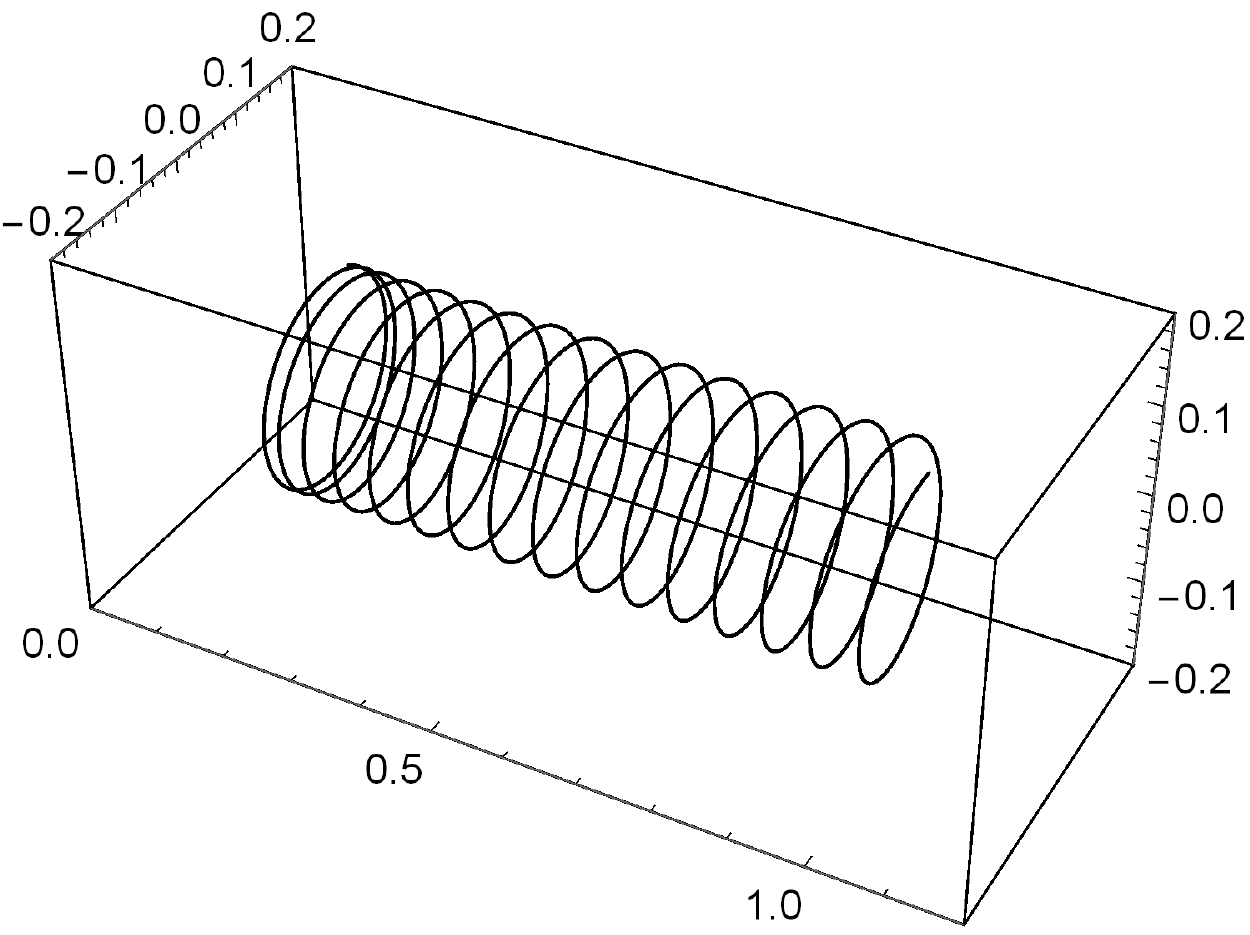}
\put(11,9){\color{Black}\vector(-1,-1){4.1}}
\put(11,9){\color{Black}\vector(3,-1){5}}
\put(11,9){\color{Black}\vector(0,4){5}}
\put (10,15) {$x$}
\put (16.5,6) {$z$}
\put (4.1,3.5) {$y$}
\end{overpic}
}\end{center} & \begin{center} \scalebox{0.6}{
\begin{overpic}[width=\columnwidth]{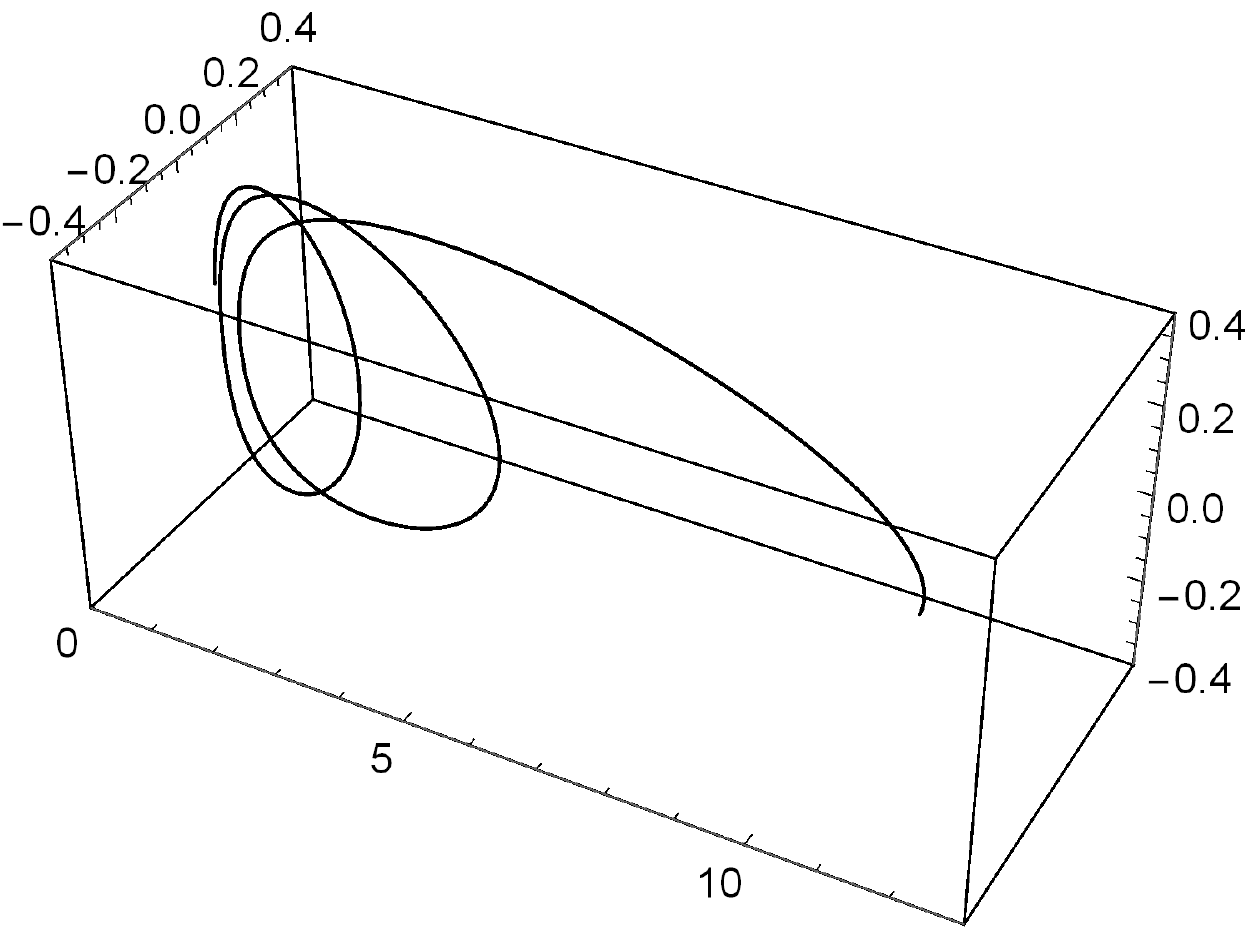}
\put(11,9){\color{Black}\vector(-1,-1){4.1}}
\put(11,9){\color{Black}\vector(3,-1){5}}
\put(11,9){\color{Black}\vector(0,4){5}}
\put (10,15) {$x$}
\put (16.5,6) {$z$}
\put (4.1,3.5) {$y$}
\end{overpic}
}\end{center} 
\\
\hline
Arrival time distributions for \(L=50\) and \(\omega=500\) & \begin{center}\begin{overpic}[width=0.6\columnwidth]{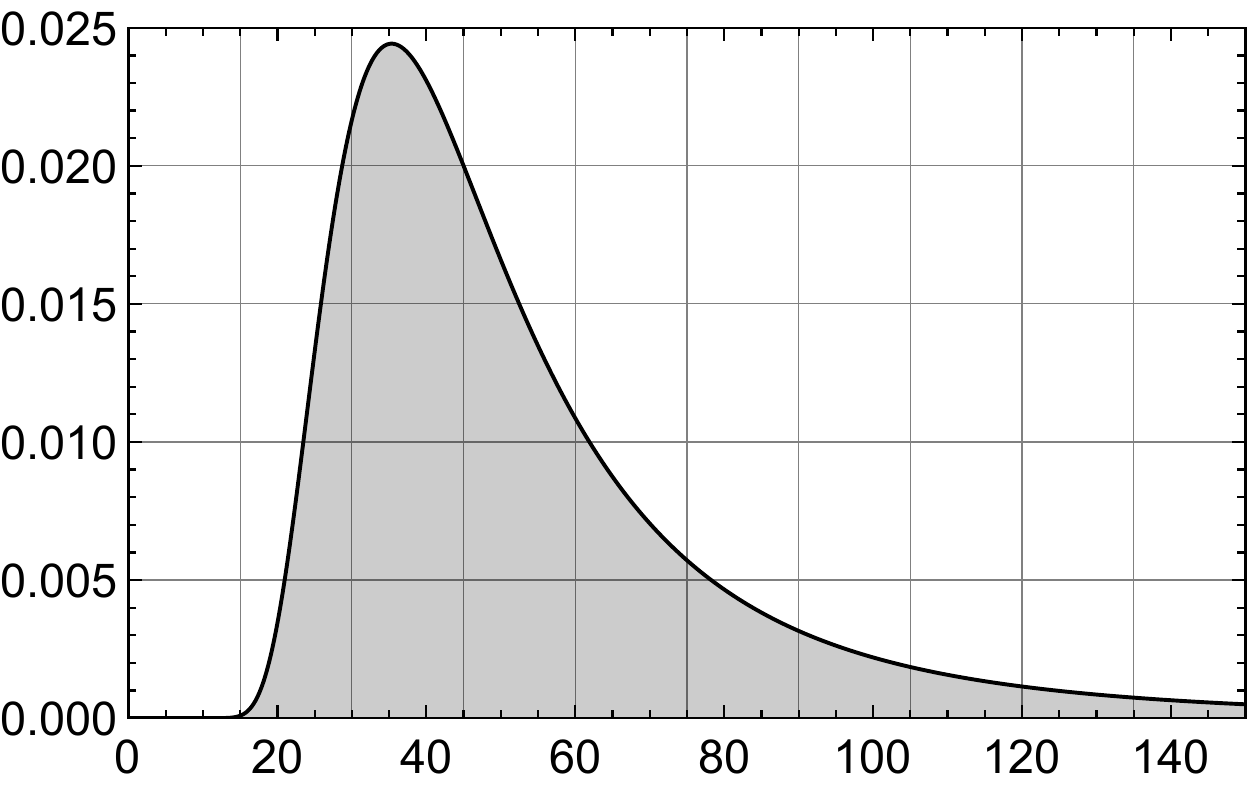}
\put (50,-3) {$\tau$}
\put (-10,32) {\rotatebox[origin=c]{90}{$\Pi_\up(\tau)$}}
\end{overpic}\end{center} & \begin{center}\begin{overpic}[width=0.6\columnwidth]{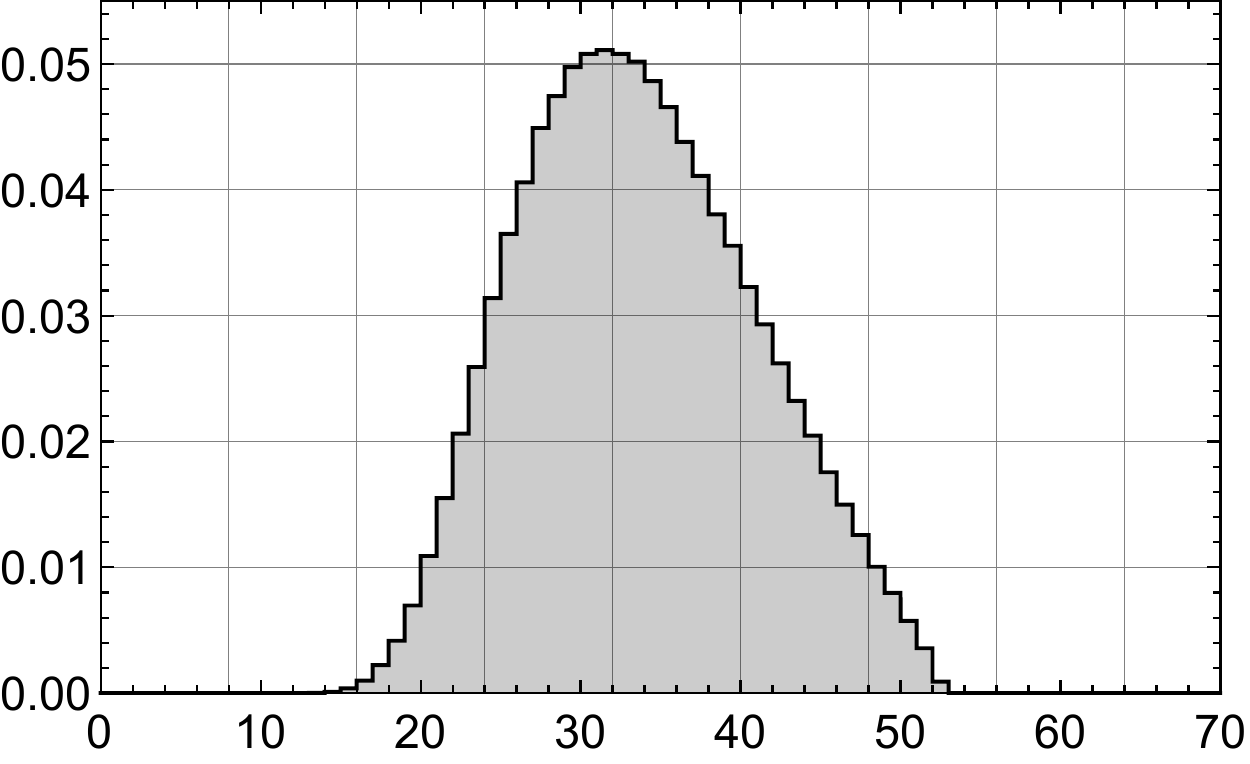}
\put (50,-3) {$\tau$}
\put (-10,32) {\rotatebox[origin=c]{90}{$\Pi_\updown(\tau)$}}
\put(81.6,14){\color{Black}\vector(-1,-1){7}}
\put (82,15.8) {$\tau_{\texttt{max}}$}
\end{overpic}\end{center}
\\
\hline
Distribution function &\begin{center}\(\displaystyle\Pi_\up (\tau)=\frac{4 L^3}{\lambda_0\sqrt{\pi}}\frac{\tau\, e^{-\frac{L^2}{1+\tau^2}}}{(1+\tau^2)^{5/2}}\)\end{center} & no closed form expression for \(\Pi_\updown(\tau)\) 
\\
\hline
Behavior for large \(\tau\)& \begin{center}
Heavy tailed \(\sim\frac{4L^3}{\lambda_0\sqrt{\pi}}\,\tau^{-4}+\mathcal{O}\big(\tau^{-6}\big)\), as \(\tau\to\infty\).     
\end{center} & \begin{center}
    Vanishes for all \(\tau>\tau_\texttt{max}.\)
\end{center}
\\
\hline
Behavior for small \(\omega\) & \begin{center}Independent of \(\omega\)\end{center} & \begin{center}
    Reduces to \(\Pi_\up(\tau)\), as \(\omega\to0\), while \(\tau_\texttt{max}\to\infty\).
\end{center}
\\
\hline
Behavior for large \(\omega\)& Independent of \(\omega\) & \begin{center}Convergence to \(\Pi_s(\tau)\), as \(\omega\to\infty\), while \(\tau_\texttt{max}\to\sqrt{L^2-1}\).\end{center}
\\
\hline 
Arrival time moments &\begin{center}\(\label{moments}
    \expval{\tau^\mu}_\up=\frac{4L^3}{3\lambda_0\sqrt{\pi}}\!\begin{cases}
       ~\,\,_1F_1\!\left(1;\frac{5}{2};-L^2\right)\!,~&\mu=1\\
       2\,_1F_1\!\left(\frac{1}{2};\frac{5}{2};-L^2\right)\!,~&\mu=2\\
       \infty,~&\mu>2
      \end{cases}
\)\end{center} & \begin{center}All moments are finite.\end{center}
\\
\hline
\end{tabularx}
    \caption{Overview of results and essential details of the paper.}
    \label{summarytab}
\end{table*}

\section*{Acknowledgments}
We thank J. M. Wilkes for critically reviewing our manuscript and suggesting numerous edits, which improved the paper significantly. We thank Dustin Lazarovici for a careful reading of a preliminary version of our paper. Thanks are also due to Matthias Lienert, Dipankar Home and Peter R. Holland for helpful discussions. M.N. acknowledges funding from the Elite Network of Bavaria, through the Junior Research Group `Interaction Between Light and Matter'.

\appendix

\section{Time evolution of \(\Psi_0\)}\label{evolpsi}
The Pauli equation \eqref{pauli} with initial condition \eqref{generic} can be solved as follows: applying the identity \((\bb{\sigma}\cdot\grad)^2=\laplacian\mathds{1}\) (where \(\mathds{1}\) is the \(2\times2\) unit matrix), the right-hand side of \eqref{pauli} becomes diagonal, essentially simplifying it to the Schr\"{o}dinger equation
\begin{equation}\label{sch}
    i\frac{\partial\psi_t}{\partial t}=-\frac{1}{2}\laplacian\psi_t+\left[\frac{\omega^2\!}{2}(x^2+y^2)+v(z)\right]\!\psi_t,
\end{equation}
for the spatial part of the spinor wave function \(\Psi\), with initial condition \eqref{spatial}. The constant spinor forming the spin part of the wave function remains unchanged. Now, employing the ansatz
\begin{equation}
    \psi_t(\bb{r})=\varphi_t(z)\kern0.03em e^{-\frac{\omega}{2}(x^2+y^2)-i\omega t}
\end{equation}
in \eqref{sch}, we arrive at the PDE
\begin{equation}\label{PDE}
    i\frac{\partial\varphi_t}{\partial t}=-\frac{1}{2}\frac{\partial^2\varphi_t}{\partial z^2}+v(z)\kern0.03em\varphi_t,
\end{equation}
for the function \(\varphi_t\), which satisfies \(\varphi_0(z)=A\,\theta(z)\,z\,e^{-\frac{z^2}{2}}\). Equation \eqref{PDE} is simply the one-dimensional Schr\"odinger equation for a particle subject to a hard-wall potential barrier at \(z=0\), thus \(\varphi_t(z)=0\) for any \(z\le0\). In the region \(z>0\), the solution of eq. \eqref{PDE} can be written as
\begin{equation}\label{symmetric}
    \varphi_t(z)=\int_0^{\infty}\!\!\!\mathrm{d}z^\prime~K\big(z,t\,|\,z^\prime,0\big)\kern0.04em\varphi_0(z^\prime),
\end{equation}
where
\begin{equation}
    K\big(z,t\,|\,z^\prime,0\big)=\frac{e^{\frac{i}{2t}(z-z^\prime)^2}}{\sqrt{2\pi it}}-\frac{e^{\frac{i}{2t}(z+z^\prime)^2}}{\sqrt{2\pi it}}
\end{equation}
is the propagator (or Green's function) of \eqref{PDE} \cite{propagator}. Exploiting the symmetry of the integrand in \eqref{symmetric} to extend the integral to \(-\infty<z^\prime<\infty\) allows writing the solution as 
\begin{align}
    \varphi_t(z)&=\frac{A}{2\sqrt{2\pi it}}\int_{-\infty}^{\infty}\!\!\!\mathrm{d}z^\prime~z^\prime\!\left[\exp\!\left(\frac{i}{2t}(z-z^\prime)^2-\frac{z^{\prime\kern0.03em2}}{2}\right)\right.\nonumber\\&\left.\kern10em-\exp\!\left(\frac{i}{2t}(z+z^\prime)^2-\frac{z^{\prime\kern0.03em2}}{2}\right)\right]\nonumber\\
    &=\frac{A\kern0.04em e^{iz^2/2t}}{\sqrt{2\pi it}}\int_{-\infty}^{\infty}\!\!\!\mathrm{d}z^\prime~z^\prime(-i)\sin\!\left(\!\frac{zz^\prime}{t}\kern-0.03em\right)\!e^{-\left(\frac{1}{2}-\frac{i}{2t}\right)z^{\prime2}}\nonumber\\
    &=\frac{A\kern0.04em e^{iz^2/2t}}{\sqrt{2\pi it}}\int_{-\infty}^{\infty}\!\!\!\mathrm{d}z^\prime~z^\prime\exp\!\left(\!-\!\left(\frac{1}{2}-\frac{i}{2t}\right)\!z^{\prime\kern0.03em 2}-\frac{iz}{t}z^\prime\right)\nonumber\\
    &=\frac{A\kern0.04em z}{(1+it)^{3/2}}\,e^{-\frac{z^2}{2(1+it)}},
\end{align}
using the identity
\begin{equation}
    \frac{2}{\sqrt{\pi}}\int_{-\infty}^{\infty}\!\!\!\mathrm{d}x~x\,e^{-ax^2+bx}=\frac{b}{a^{3/2}}\kern0.03em e^{b^2/4a},\quad\mathrm{Re}[a]>0.
\end{equation}
The final solution thus reduces to \eqref{spatialt}.

\section{Distribution of \(t_s\)}\label{cherish}
Equation \eqref{tbounds} expresses \(t_s\) as a function of \(\xi_b\), which takes values in the interval \([1,\infty)\) (see eq. \eqref{important}). Thus, the distribution of \(t_s\) may be written as
\begin{align}\label{ddd}
    \Pi_s(t)&=\int_1^{\infty}\!\!\!\mathrm{d}\xi_b~\delta\big(t_s(\xi_b)-t\big)\kern0.05em\Lambda(\xi_b),
\end{align}
where \(\Lambda\) is the distribution of \(\xi_b\), given by
\begin{align}\label{distxi}
   \Lambda(\xi_b)=1/\lambda_0\kern-1.3em\int\limits_{0<Z_0<L}\kern-1em\mathrm{d}^3\bb{R}_0~\delta\big(\xi_b(Y_0,Z_0)-\xi_b\big)|\Psi_0|^2(\bb{R}_0).
\end{align}
Here, \(\xi_b(Y_0,Z_0)\) is defined via eq. \eqref{roots} and \eqref{g}:
\begin{equation}\label{nnn}
    \xi_b(Y_0,Z_0)\equiv\xi_b(g(Y_0,Z_0))=\sqrt{-W_{-1}\big(\!-Z_0^2e^{-Z_0^2-\omega Y_0^2}\big)}.
\end{equation}
Substituting the definition of \(t_s\), eq. \eqref{tbounds} in \eqref{ddd}, we obtain
\begin{align}
    \kern-0.3em\Pi_s(t)&=\!\int_1^L\!\!\!\mathrm{d}\xi_b~\delta\!\left(\sqrt{\frac{L^2}{\xi_b^2}-1}-t\right)\Lambda(\xi_b)\nonumber\\&\kern8em+\,\delta\big(t\big)\!\!\int_L^{\infty}\!\!\!\mathrm{d}\xi_b~\Lambda(\xi_b).\label{see it}
\end{align}
We shall denote the integral multiplying $\delta(t)$ by
\begin{equation}
    \eta\coloneqq\int_L^{\infty}\!\!\!\mathrm{d}\xi_b~\Lambda(\xi_b).
\end{equation}
The remaining integral in \eqref{see it} can be evaluated with the help of identity \eqref{delta}, exactly as in Section \ref{UPArrival}, with the final result:
\begin{align}\label{llimdist}
    \Pi_s(t)=\frac{t\kern0.03em L}{(1+t^2)^{3/2}}\,\theta\big(\sqrt{L^2-1}-t\big)\,\Lambda\!\left(\frac{L}{\sqrt{1+t^2}}\right)+\eta\,\delta(t).
\end{align}
Note that $\Pi_s(t)$ vanishes for any $t>\sqrt{L^2-1}$, regardless of the specific form of \(\Lambda(\xi_b)\).

Next, we turn to the evaluation of \(\Lambda(\xi_b)\). Substituting $|\Psi_0|^2(\bb{R}_0)$ (eq. \eqref{indist}) in \eqref{distxi}, and integrating over $X_0$, yields
\begin{equation}
    \!\!\Lambda(\xi_b)=\frac{4\sqrt{\omega}}{\pi\,\lambda_0}\!\int_0^L\!\!\!\mathrm{d}Z_0~Z_0^2\!\int_{-\infty}^{\infty}\!\!\!\mathrm{d}Y_0~\delta\big(\xi_b(Y_0,Z_0)-\xi_b\big)e^{-Z_0^2-\omega Y_0^2}.\label{interim}
\end{equation}
Once again, recalling identity \eqref{delta}, with
\begin{equation}
    \phi(Y_0)=\xi_b(Y_0,Z_0)-\xi_b,
\end{equation}
we compute the zeros of $\phi$, satisfying $\xi_b(Y_0,Z_0)=\xi_b$,
\begin{align}
    &\Rightarrow W_{-1}\big(-Z_0^2e^{-Z_0^2-\omega Y_0^2}\big)=-\xi_b^2\nonumber\\
    &\Rightarrow e^{-\omega Y_0^2}=\frac{\xi_b^2\,e^{-\xi_b^2}}{Z_0^2\,e^{-Z_0^2}}\label{step1}\\
    &\Rightarrow Y_0=\pm\sqrt{\frac{2\ln(Z_0/\xi_b)+\xi_b^2-Z_0^2}{\omega}}\equiv Y_{0\pm}.\label{Ystar}
\end{align}
In \eqref{step1}, we invoked the defining property of the Lambert W function: \(W(a)\!=\!b\Leftrightarrow a\!=\!be^b\) \cite{Knuth}. We evaluate \(\phi'(Y_{0\pm})\) as follows:
\begin{align}
    \phi'(Y_{0\pm})&=\left.\frac{\partial\xi_b(Y_0,Z_0)}{\partial Y_0}\right|_{Y_{0\pm}}=\left.\frac{\partial\xi_b(g)}{\partial g}\frac{\partial g}{\partial Y_0}\right|_{Y_{0\pm}}\nonumber\\
    &=\left.\frac{-W_{-1}(g)}{2g\xi_b(g)(1+W_{-1}(g))}\times(-2g\omega Y_0)\right|_{Y_{0\pm}}\nonumber\\
    &=\omega\,Y_{0\pm}\kern0.03em\frac{\xi_b}{\xi_b^2-1}.
\end{align}
Here, we used the identity \(W'=W/z(1+W)\) \cite{Knuth}. Putting all the pieces together yields
\begin{align}
    \delta\big(\xi_b(Y_0,Z_0)-\xi_b\big)&=\theta\!\left(Z_0^2\,e^{-Z_0^2}-\xi_b^2\,e^{-\xi_b^2}\right)\frac{\xi_b^2-1}{\omega\,\xi_b Y_{0+}}\nonumber\\
    &\quad\times\big[\delta(Y_0-Y_{0+})+\delta(Y_0-Y_{0-})\big],\label{deltafappendix}
\end{align}
via \eqref{delta}. Note that the Heaviside function \(\theta(\cdot)\) eliminates any \(Z_0\) that gives rise to an imaginary \(Y_{0\pm}\), which therefore does not contribute to the integral \eqref{interim}. Substituting \eqref{deltafappendix} into \eqref{interim} and evaluating the integral over \(Y_0\) yields
\begin{align}
    \Lambda(\xi_b)&=\frac{8\kern0.03em(\xi_b^2-1)}{\pi\,\lambda_0\,\xi_b\sqrt{\omega}}\!\int_0^L\!\!\!\mathrm{d}Z_0~\frac{Z_0^2}{\,Y_{0+}^2\!}\theta\!\left(Z_0^2\,e^{-Z_0^2}-\xi_b^2\,e^{-\xi_b^2}\right)\nonumber\\[-5pt]
    &\kern6cm\times e^{-Z_0^2-\omega Y_{0+}^2}\nonumber\\
   &=\frac{8\kern0.07em\xi_b}{\pi\kern0.07em\lambda_0}\kern0.04em(\xi_b^2-1)\kern0.03em e^{-\xi_b^2}\!\!\bigintssss_0^L\!\!\!\mathrm{d}Z_0\frac{\theta\!\left(Z_0^2\,e^{-Z_0^2}-\xi_b^2\,e^{-\xi_b^2}\right)}{\sqrt{2\ln(Z_0/\xi_b)+\xi_b^2-Z_0^2}},\label{finally}
\end{align}
using eq. \eqref{step1} and \eqref{Ystar}. Note that \(\omega\) dropped out obligingly in the previous step. Now, for a given \(\xi_b\ge-1\), the inequality \(Z_0^2\,e^{-Z_0^2}>\xi_b^2\,e^{-\xi_b^2}\) implies
\begin{equation*}
    \underbrace{\sqrt{-W_0\big(-\xi_b^2e^{-\xi_b^2}\big)}}_{\displaystyle\,\,\eqqcolon\ell(\xi_b)}<Z_0<\underbrace{\sqrt{-W_{-1}\big(-\xi_b^2e^{-\xi_b^2}\big)}}_{\displaystyle=\xi_b}\,,
\end{equation*}
which incorporated into \eqref{finally}, yields the final result
\begin{align}
   \Lambda(\xi_b)=\frac{8\kern0.07em\xi_b}{\pi\kern0.07em\lambda_0}\kern0.04em(\xi_b^2-1)\kern0.03em e^{-\xi_b^2}\!\!\bigintssss_{\ell(\xi_b)}^{\,\min\{\xi_b,L\}}\!\!\!\!\!\!\frac{\mathrm{d}Z_0}{\sqrt{2\ln(Z_0/\xi_b)+\xi_b^2-Z_0^2}}.
\end{align}
Since we evaluate \(\Lambda(\cdot)\) at \(L/\sqrt{1+t^2}\) in \eqref{llimdist}, the upper limit of the integral can be simply replaced by \(\xi_b\).

\bibliography{sample}

\begin{thebibliography}{38}%
\makeatletter
\providecommand \@ifxundefined [1]{%
 \@ifx{#1\undefined}
}%
\providecommand \@ifnum [1]{%
 \ifnum #1\expandafter \@firstoftwo
 \else \expandafter \@secondoftwo
 \fi
}%
\providecommand \@ifx [1]{%
 \ifx #1\expandafter \@firstoftwo
 \else \expandafter \@secondoftwo
 \fi
}%
\providecommand \natexlab [1]{#1}%
\providecommand \enquote  [1]{``#1''}%
\providecommand \bibnamefont  [1]{#1}%
\providecommand \bibfnamefont [1]{#1}%
\providecommand \citenamefont [1]{#1}%
\providecommand \href@noop [0]{\@secondoftwo}%
\providecommand \href [0]{\begingroup \@sanitize@url \@href}%
\providecommand \@href[1]{\@@startlink{#1}\@@href}%
\providecommand \@@href[1]{\endgroup#1\@@endlink}%
\providecommand \@sanitize@url [0]{\catcode `\\12\catcode `\$12\catcode
  `\&12\catcode `\#12\catcode `\^12\catcode `\_12\catcode `\%12\relax}%
\providecommand \@@startlink[1]{}%
\providecommand \@@endlink[0]{}%
\providecommand \url  [0]{\begingroup\@sanitize@url \@url }%
\providecommand \@url [1]{\endgroup\@href {#1}{\urlprefix }}%
\providecommand \urlprefix  [0]{URL }%
\providecommand \Eprint [0]{\href }%
\providecommand \doibase [0]{http://dx.doi.org/}%
\providecommand \selectlanguage [0]{\@gobble}%
\providecommand \bibinfo  [0]{\@secondoftwo}%
\providecommand \bibfield  [0]{\@secondoftwo}%
\providecommand \translation [1]{[#1]}%
\providecommand \BibitemOpen [0]{}%
\providecommand \bibitemStop [0]{}%
\providecommand \bibitemNoStop [0]{.\EOS\space}%
\providecommand \EOS [0]{\spacefactor3000\relax}%
\providecommand \BibitemShut  [1]{\csname bibitem#1\endcsname}%
\let\auto@bib@innerbib\@empty
\bibitem [{\citenamefont {Muga}\ \emph {et~al.}(2008)\citenamefont {Muga},
  \citenamefont {Mayato},\ and\ \citenamefont {Egusquiza}}]{MUGA}%
  \BibitemOpen
  \bibinfo {editor} {\bibfnamefont {J.~G.}\ \bibnamefont {Muga}}, \bibinfo
  {editor} {\bibfnamefont {R.~S.}\ \bibnamefont {Mayato}}, \ and\ \bibinfo
  {editor} {\bibfnamefont {{\'{I}}.~L.}\ \bibnamefont {Egusquiza}},\ eds.,\
  \href {\doibase 10.1007/978-3-540-73473-4} {\emph {\bibinfo {title} {Time in
  Quantum Mechanics}}},\ \bibinfo {edition} {2nd}\ ed.,\ \bibinfo {series}
  {Lect. Notes Phys. 734}, Vol.~\bibinfo {volume} {1}\ (\bibinfo  {publisher}
  {Springer},\ \bibinfo {address} {Berlin Heidelberg},\ \bibinfo {year}
  {2008})\BibitemShut {NoStop}%
\bibitem [{\citenamefont {Muga}\ and\ \citenamefont {Leavens}(2000)}]{MUGA1}%
  \BibitemOpen
  \bibfield  {author} {\bibinfo {author} {\bibfnamefont {J.~G.}\ \bibnamefont
  {Muga}}\ and\ \bibinfo {author} {\bibfnamefont {C.~R.}\ \bibnamefont
  {Leavens}},\ }\href {\doibase 10.1016/S0370-1573(00)00047-8} {\bibfield
  {journal} {\bibinfo  {journal} {Phys. Rep.}\ }\textbf {\bibinfo {volume}
  {338}},\ \bibinfo {pages} {353} (\bibinfo {year} {2000})}\BibitemShut
  {NoStop}%
\bibitem [{\citenamefont {Allcock}(1969{\natexlab{a}})}]{Allcock1}%
  \BibitemOpen
  \bibfield  {author} {\bibinfo {author} {\bibfnamefont {G.~R.}\ \bibnamefont
  {Allcock}},\ }\href {\doibase 10.1016/0003-4916(69)90251-6} {\bibfield
  {journal} {\bibinfo  {journal} {Ann. Phys.}\ }\textbf {\bibinfo {volume}
  {53}},\ \bibinfo {pages} {253} (\bibinfo {year}
  {1969}{\natexlab{a}})}\BibitemShut {NoStop}%
\bibitem [{\citenamefont {Blanchard}\ and\ \citenamefont
  {Fr{\"{o}}hlich}(2015)}]{Vona}%
  \BibitemOpen
  \bibinfo {editor} {\bibfnamefont {P.}~\bibnamefont {Blanchard}}\ and\
  \bibinfo {editor} {\bibfnamefont {J.}~\bibnamefont {Fr{\"{o}}hlich}},\ eds.,\
  \href {\doibase 10.1007/978-3-662-46422-9} {\emph {\bibinfo {title} {The
  Message of Quantum Science: Attempts Towards a Synthesis}}},\ Lect. Notes
  Phys. 899\ (\bibinfo  {publisher} {Springer},\ \bibinfo {address} {Berlin
  Heidelberg},\ \bibinfo {year} {2015})\ Chap.~\bibinfo {chapter}
  {5}\BibitemShut {NoStop}%
\bibitem [{\citenamefont {Aharonov}\ and\ \citenamefont {Bohm}(1961)}]{AhBohm}%
  \BibitemOpen
  \bibfield  {author} {\bibinfo {author} {\bibfnamefont {Y.}~\bibnamefont
  {Aharonov}}\ and\ \bibinfo {author} {\bibfnamefont {D.}~\bibnamefont
  {Bohm}},\ }\href {\doibase 10.1103/PhysRev.122.1649} {\bibfield  {journal}
  {\bibinfo  {journal} {Phys. Rev.}\ }\textbf {\bibinfo {volume} {122}},\
  \bibinfo {pages} {1649} (\bibinfo {year} {1961})}\BibitemShut {NoStop}%
\bibitem [{\citenamefont {Leavens}(2002)}]{Leavens1}%
  \BibitemOpen
  \bibfield  {author} {\bibinfo {author} {\bibfnamefont {C.~R.}\ \bibnamefont
  {Leavens}},\ }\href {\doibase 10.1016/S0375-9601(02)01239-2} {\bibfield
  {journal} {\bibinfo  {journal} {Physics Letters A}\ }\textbf {\bibinfo
  {volume} {303}},\ \bibinfo {pages} {154} (\bibinfo {year}
  {2002})}\BibitemShut {NoStop}%
\bibitem [{\citenamefont {Mielnik}\ and\ \citenamefont
  {Torres-Vega}(2005)}]{Mielnik}%
  \BibitemOpen
  \bibfield  {author} {\bibinfo {author} {\bibfnamefont {B.}~\bibnamefont
  {Mielnik}}\ and\ \bibinfo {author} {\bibfnamefont {G.}~\bibnamefont
  {Torres-Vega}},\ }\href@noop {} {\bibfield  {journal} {\bibinfo  {journal}
  {Concepts of Physics.}\ }\textbf {\bibinfo {volume} {II}},\ \bibinfo {pages}
  {81} (\bibinfo {year} {2005})}\BibitemShut {NoStop}%
\bibitem [{\citenamefont {Galapon}\ \emph {et~al.}(2004)\citenamefont
  {Galapon}, \citenamefont {Caballar},\ and\ \citenamefont {Jr}}]{CTOA}%
  \BibitemOpen
  \bibfield  {author} {\bibinfo {author} {\bibfnamefont {E.~A.}\ \bibnamefont
  {Galapon}}, \bibinfo {author} {\bibfnamefont {R.~F.}\ \bibnamefont
  {Caballar}}, \ and\ \bibinfo {author} {\bibfnamefont {R.~T.~B.}\ \bibnamefont
  {Jr}},\ }\href {\doibase 10.1103/PhysRevLett.93.180406} {\bibfield  {journal}
  {\bibinfo  {journal} {Phys. Rev. Lett.}\ }\textbf {\bibinfo {volume} {93}},\
  \bibinfo {pages} {180406} (\bibinfo {year} {2004})}\BibitemShut {NoStop}%
\bibitem [{\citenamefont {D{\"{u}}rr}\ \emph {et~al.}(2004)\citenamefont
  {D{\"{u}}rr}, \citenamefont {Goldstein},\ and\ \citenamefont
  {Zangh{\`{I}}}}]{DGZOperators}%
  \BibitemOpen
  \bibfield  {author} {\bibinfo {author} {\bibfnamefont {D.}~\bibnamefont
  {D{\"{u}}rr}}, \bibinfo {author} {\bibfnamefont {S.}~\bibnamefont
  {Goldstein}}, \ and\ \bibinfo {author} {\bibfnamefont {N.}~\bibnamefont
  {Zangh{\`{I}}}},\ }\href {\doibase 10.1023/B:JOSS.0000} {\bibfield  {journal}
  {\bibinfo  {journal} {J. Stat. Phys.}\ }\textbf {\bibinfo {volume} {116}},\
  \bibinfo {pages} {959} (\bibinfo {year} {2004})}\BibitemShut {NoStop}%
\bibitem [{\citenamefont {Kijowski}(1974)}]{KijowskiPOVM}%
  \BibitemOpen
  \bibfield  {author} {\bibinfo {author} {\bibfnamefont {J.}~\bibnamefont
  {Kijowski}},\ }\href {\doibase 10.1016/S0034-4877(74)80004-2} {\bibfield
  {journal} {\bibinfo  {journal} {Reports on Mathematical Physics}\ }\textbf
  {\bibinfo {volume} {6}},\ \bibinfo {pages} {361} (\bibinfo {year}
  {1974})}\BibitemShut {NoStop}%
\bibitem [{\citenamefont {Werner}(1986)}]{Werner}%
  \BibitemOpen
  \bibfield  {author} {\bibinfo {author} {\bibfnamefont {R.}~\bibnamefont
  {Werner}},\ }\href {\doibase 10.1063/1.527184} {\bibfield  {journal}
  {\bibinfo  {journal} {Journal of Mathematical Physics}\ }\textbf {\bibinfo
  {volume} {27}},\ \bibinfo {pages} {793} (\bibinfo {year} {1986})}\BibitemShut
  {NoStop}%
\bibitem [{\citenamefont {Anastopoulos}\ and\ \citenamefont
  {Savvidou}(2006)}]{AS}%
  \BibitemOpen
  \bibfield  {author} {\bibinfo {author} {\bibfnamefont {C.}~\bibnamefont
  {Anastopoulos}}\ and\ \bibinfo {author} {\bibfnamefont {N.}~\bibnamefont
  {Savvidou}},\ }\href {\doibase 10.1063/1.2399085} {\bibfield  {journal}
  {\bibinfo  {journal} {Journal of Mathematical Physics}\ }\textbf {\bibinfo
  {volume} {47}},\ \bibinfo {pages} {122106} (\bibinfo {year}
  {2006})}\BibitemShut {NoStop}%
\bibitem [{\citenamefont {{Tumulka}}(2016)}]{Rodi}%
  \BibitemOpen
  \bibfield  {author} {\bibinfo {author} {\bibfnamefont {R.}~\bibnamefont
  {{Tumulka}}},\ }\href@noop {} {\bibfield  {journal} {\bibinfo  {journal}
  {ArXiv e-prints}\ } (\bibinfo {year} {2016})},\ \Eprint
  {http://arxiv.org/abs/1601.03715} {arXiv:1601.03715} \BibitemShut {NoStop}%
\bibitem [{\citenamefont {Das}\ \emph {et~al.}()\citenamefont {Das},
  \citenamefont {N{\"{o}}th},\ and\ \citenamefont {D{\"{u}}rr}}]{complexPot}%
  \BibitemOpen
  \bibfield  {author} {\bibinfo {author} {\bibfnamefont {S.}~\bibnamefont
  {Das}}, \bibinfo {author} {\bibfnamefont {M.}~\bibnamefont {N{\"{o}}th}}, \
  and\ \bibinfo {author} {\bibfnamefont {D.}~\bibnamefont {D{\"{u}}rr}},\
  }\href@noop {} {\enquote {\bibinfo {title} {Exotic {B}ohmian arrival times of
  spin-1/2 particles {II}–-{E}xperimental considerations},}\ }\bibinfo {note}
  {{i}n preparation}\BibitemShut {NoStop}%
\bibitem [{\citenamefont {Allcock}(1969{\natexlab{b}})}]{Allcock2}%
  \BibitemOpen
  \bibfield  {author} {\bibinfo {author} {\bibfnamefont {G.~R.}\ \bibnamefont
  {Allcock}},\ }\href {\doibase 10.1016/0003-4916(69)90252-8} {\bibfield
  {journal} {\bibinfo  {journal} {Ann. Phys.}\ }\textbf {\bibinfo {volume}
  {53}},\ \bibinfo {pages} {286} (\bibinfo {year}
  {1969}{\natexlab{b}})}\BibitemShut {NoStop}%
\bibitem [{\citenamefont {Kellers}(2017)}]{LeoThesis}%
  \BibitemOpen
  \bibfield  {author} {\bibinfo {author} {\bibfnamefont {L.}~\bibnamefont
  {Kellers}},\ }\emph {\bibinfo {title} {Select Trajectory Methods as a Tool in
  Numerical Quantum Mechanics}},\ \href@noop {} {Master's thesis},\ \bibinfo
  {school} {LMU Munich {\&} TU Munich} (\bibinfo {year} {2017}),\ \bibinfo
  {note}
  {\url{www.mathematik.uni-muenchen.de/\%7Ebohmmech/theses/Kellers_Leopold_MA.pdf}}\BibitemShut
  {NoStop}%
\bibitem [{\citenamefont {Hils}\ \emph {et~al.}(1998)\citenamefont {Hils},
  \citenamefont {Felber}, \citenamefont {G\"ahler}, \citenamefont {Gl\"aser},
  \citenamefont {Golub}, \citenamefont {Habicht},\ and\ \citenamefont
  {Wille}}]{Hils}%
  \BibitemOpen
  \bibfield  {author} {\bibinfo {author} {\bibfnamefont {T.}~\bibnamefont
  {Hils}}, \bibinfo {author} {\bibfnamefont {J.}~\bibnamefont {Felber}},
  \bibinfo {author} {\bibfnamefont {R.}~\bibnamefont {G\"ahler}}, \bibinfo
  {author} {\bibfnamefont {W.}~\bibnamefont {Gl\"aser}}, \bibinfo {author}
  {\bibfnamefont {R.}~\bibnamefont {Golub}}, \bibinfo {author} {\bibfnamefont
  {K.}~\bibnamefont {Habicht}}, \ and\ \bibinfo {author} {\bibfnamefont
  {P.}~\bibnamefont {Wille}},\ }\href {\doibase 10.1103/PhysRevA.58.4784}
  {\bibfield  {journal} {\bibinfo  {journal} {Phys. Rev. A}\ }\textbf {\bibinfo
  {volume} {58}},\ \bibinfo {pages} {4784} (\bibinfo {year}
  {1998})}\BibitemShut {NoStop}%
\bibitem [{\citenamefont {Uehara}\ \emph {et~al.}(1990)\citenamefont {Uehara}
  \emph {et~al.}}]{NICE}%
  \BibitemOpen
  \bibfield  {author} {\bibinfo {author} {\bibfnamefont {Y.}~\bibnamefont
  {Uehara}} \emph {et~al.},\ }\href {\doibase 10.1143/JJAP.29.2858} {\bibfield
  {journal} {\bibinfo  {journal} {Jpn. J. Appl. Phys}\ }\textbf {\bibinfo
  {volume} {29}},\ \bibinfo {pages} {2858} (\bibinfo {year}
  {1990})}\BibitemShut {NoStop}%
\bibitem [{\citenamefont {Kothe}\ \emph {et~al.}(2013)\citenamefont {Kothe}
  \emph {et~al.}}]{Kothe}%
  \BibitemOpen
  \bibfield  {author} {\bibinfo {author} {\bibfnamefont {A.}~\bibnamefont
  {Kothe}} \emph {et~al.},\ }\href {\doibase 10.1063/1.4791792} {\bibfield
  {journal} {\bibinfo  {journal} {Review of Scientific Instruments}\ }\textbf
  {\bibinfo {volume} {84}},\ \bibinfo {pages} {023106} (\bibinfo {year}
  {2013})}\BibitemShut {NoStop}%
\bibitem [{\citenamefont {Groot-Berning}\ \emph {et~al.}(2018)\citenamefont
  {Groot-Berning} \emph {et~al.}}]{Kaler}%
  \BibitemOpen
  \bibfield  {author} {\bibinfo {author} {\bibfnamefont {k.}~\bibnamefont
  {Groot-Berning}} \emph {et~al.},\ }\href@noop {} {\bibfield  {journal}
  {\bibinfo  {journal} {ArXiv e-prints}\ } (\bibinfo {year} {2018})},\ \Eprint
  {http://arxiv.org/abs/1807.05975} {arXiv:1807.05975} \BibitemShut {NoStop}%
\bibitem [{\citenamefont {Szriftgiser}\ \emph {et~al.}(1996)\citenamefont
  {Szriftgiser}, \citenamefont {Gu\'ery-Odelin}, \citenamefont {Arndt},\ and\
  \citenamefont {Dalibard}}]{Dalibard}%
  \BibitemOpen
  \bibfield  {author} {\bibinfo {author} {\bibfnamefont {P.}~\bibnamefont
  {Szriftgiser}}, \bibinfo {author} {\bibfnamefont {D.}~\bibnamefont
  {Gu\'ery-Odelin}}, \bibinfo {author} {\bibfnamefont {M.}~\bibnamefont
  {Arndt}}, \ and\ \bibinfo {author} {\bibfnamefont {J.}~\bibnamefont
  {Dalibard}},\ }\href {\doibase 10.1103/PhysRevLett.77.4} {\bibfield
  {journal} {\bibinfo  {journal} {Phys. Rev. Lett.}\ }\textbf {\bibinfo
  {volume} {77}},\ \bibinfo {pages} {4} (\bibinfo {year} {1996})}\BibitemShut
  {NoStop}%
\bibitem [{\citenamefont {Salomon}\ \emph {et~al.}(1990)\citenamefont
  {Salomon}, \citenamefont {Dalibard}, \citenamefont {Phillips}, \citenamefont
  {Clairon},\ and\ \citenamefont {Guellati}}]{Salomon}%
  \BibitemOpen
  \bibfield  {author} {\bibinfo {author} {\bibfnamefont {C.}~\bibnamefont
  {Salomon}}, \bibinfo {author} {\bibfnamefont {J.}~\bibnamefont {Dalibard}},
  \bibinfo {author} {\bibfnamefont {W.~D.}\ \bibnamefont {Phillips}}, \bibinfo
  {author} {\bibfnamefont {A.}~\bibnamefont {Clairon}}, \ and\ \bibinfo
  {author} {\bibfnamefont {S.}~\bibnamefont {Guellati}},\ }\href {\doibase
  10.1209/0295-5075/12/8/003} {\bibfield  {journal} {\bibinfo  {journal} {EPL}\
  }\textbf {\bibinfo {volume} {12}},\ \bibinfo {pages} {683} (\bibinfo {year}
  {1990})}\BibitemShut {NoStop}%
\bibitem [{\citenamefont {Fuhrmanek}\ \emph {et~al.}(2010)\citenamefont
  {Fuhrmanek}, \citenamefont {Lance}, \citenamefont {Tuchendler}, \citenamefont
  {Grangier}, \citenamefont {Sortais},\ and\ \citenamefont {Browaeys}}]{RB}%
  \BibitemOpen
  \bibfield  {author} {\bibinfo {author} {\bibfnamefont {A.}~\bibnamefont
  {Fuhrmanek}}, \bibinfo {author} {\bibfnamefont {A.~M.}\ \bibnamefont
  {Lance}}, \bibinfo {author} {\bibfnamefont {C.}~\bibnamefont {Tuchendler}},
  \bibinfo {author} {\bibfnamefont {P.}~\bibnamefont {Grangier}}, \bibinfo
  {author} {\bibfnamefont {Y.~R.~P.}\ \bibnamefont {Sortais}}, \ and\ \bibinfo
  {author} {\bibfnamefont {A.}~\bibnamefont {Browaeys}},\ }\href {\doibase
  10.1088/1367-2630/12/5/053028} {\bibfield  {journal} {\bibinfo  {journal}
  {New Journal of Physics}\ }\textbf {\bibinfo {volume} {12}},\ \bibinfo
  {pages} {053028} (\bibinfo {year} {2010})}\BibitemShut {NoStop}%
\bibitem [{\citenamefont {Delgado}\ \emph {et~al.}(2006)\citenamefont
  {Delgado}, \citenamefont {Muga},\ and\ \citenamefont
  {Garc\'{\i}a-Calder\'on}}]{ZenoMuga}%
  \BibitemOpen
  \bibfield  {author} {\bibinfo {author} {\bibfnamefont {F.}~\bibnamefont
  {Delgado}}, \bibinfo {author} {\bibfnamefont {J.~G.}\ \bibnamefont {Muga}}, \
  and\ \bibinfo {author} {\bibfnamefont {G.}~\bibnamefont
  {Garc\'{\i}a-Calder\'on}},\ }\href {\doibase 10.1103/PhysRevA.74.062102}
  {\bibfield  {journal} {\bibinfo  {journal} {Phys. Rev. A}\ }\textbf {\bibinfo
  {volume} {74}},\ \bibinfo {pages} {062102} (\bibinfo {year}
  {2006})}\BibitemShut {NoStop}%
\bibitem [{\citenamefont {Das}\ and\ \citenamefont {D{\"{u}}rr}(2019)}]{DD}%
  \BibitemOpen
  \bibfield  {author} {\bibinfo {author} {\bibfnamefont {S.}~\bibnamefont
  {Das}}\ and\ \bibinfo {author} {\bibfnamefont {D.}~\bibnamefont
  {D{\"{u}}rr}},\ }\href@noop {} {\bibfield  {journal} {\bibinfo  {journal}
  {Scientific Reports}\ } (\bibinfo {year} {2019})},\ \bibinfo {note} {in
  press},\ \Eprint {http://arxiv.org/abs/1802.07141} {arXiv:1802.07141}
  \BibitemShut {NoStop}%
\bibitem [{\citenamefont {Das}(2017)}]{SDThesis}%
  \BibitemOpen
  \bibfield  {author} {\bibinfo {author} {\bibfnamefont {S.}~\bibnamefont
  {Das}},\ }\emph {\bibinfo {title} {Arrival Time Distributions of Spin-1/2
  Particles}},\ \href@noop {} {Master's thesis},\ \bibinfo  {school} {LMU
  Munich {\&} TU Munich} (\bibinfo {year} {2017}),\ \bibinfo {note}
  {\url{http://www.mathematik.uni-muenchen.de/~bohmmech/theses/Das_Siddhant_MA.pdf}}\BibitemShut
  {NoStop}%
\bibitem [{\citenamefont {Das}\ and\ \citenamefont {D{\"{u}}rr}()}]{ddprep}%
  \BibitemOpen
  \bibfield  {author} {\bibinfo {author} {\bibfnamefont {S.}~\bibnamefont
  {Das}}\ and\ \bibinfo {author} {\bibfnamefont {D.}~\bibnamefont
  {D{\"{u}}rr}},\ }\href@noop {} {\enquote {\bibinfo {title} {Arrival time
  distributions and spin in quantum mechanics--{A} {B}ohmian perspective},}\
  }\bibinfo {note} {{i}n preparation}\BibitemShut {NoStop}%
\bibitem [{\citenamefont {Holland}\ and\ \citenamefont
  {Philippidis}(2003)}]{HollandPhilippidis}%
  \BibitemOpen
  \bibfield  {author} {\bibinfo {author} {\bibfnamefont {P.~R.}\ \bibnamefont
  {Holland}}\ and\ \bibinfo {author} {\bibfnamefont {C.}~\bibnamefont
  {Philippidis}},\ }\href {\doibase 10.1103/PhysRevA.67.062105} {\bibfield
  {journal} {\bibinfo  {journal} {Phy. Rev. A}\ }\textbf {\bibinfo {volume}
  {67}},\ \bibinfo {pages} {062105} (\bibinfo {year} {2003})}\BibitemShut
  {NoStop}%
\bibitem [{\citenamefont {Bohm}\ and\ \citenamefont {Hiley}(1993)}]{BohmHiley}%
  \BibitemOpen
  \bibfield  {author} {\bibinfo {author} {\bibfnamefont {D.}~\bibnamefont
  {Bohm}}\ and\ \bibinfo {author} {\bibfnamefont {B.~J.}\ \bibnamefont
  {Hiley}},\ }\href@noop {} {\emph {\bibinfo {title} {The Undivided Universe:
  An Ontological Interpretation of Quantum Theory}}}\ (\bibinfo  {publisher}
  {Routledge},\ \bibinfo {address} {London and New York},\ \bibinfo {year}
  {1993})\BibitemShut {NoStop}%
\bibitem [{\citenamefont {Berndl}\ \emph {et~al.}(1995)\citenamefont {Berndl},
  \citenamefont {D{\"{u}}rr}, \citenamefont {Goldstein}, \citenamefont
  {Peruzzi},\ and\ \citenamefont {Zangh{\`{I}}}}]{Berndl}%
  \BibitemOpen
  \bibfield  {author} {\bibinfo {author} {\bibfnamefont {K.}~\bibnamefont
  {Berndl}}, \bibinfo {author} {\bibfnamefont {D.}~\bibnamefont {D{\"{u}}rr}},
  \bibinfo {author} {\bibfnamefont {S.}~\bibnamefont {Goldstein}}, \bibinfo
  {author} {\bibfnamefont {G.}~\bibnamefont {Peruzzi}}, \ and\ \bibinfo
  {author} {\bibfnamefont {N.}~\bibnamefont {Zangh{\`{I}}}},\ }\href {\doibase
  10.1007/BF02101660} {\bibfield  {journal} {\bibinfo  {journal} {Commun. Math.
  Phys.}\ }\textbf {\bibinfo {volume} {173}},\ \bibinfo {pages} {647} (\bibinfo
  {year} {1995})}\BibitemShut {NoStop}%
\bibitem [{\citenamefont {Teufel}\ and\ \citenamefont
  {Tumulka}(2005)}]{RodiTufel}%
  \BibitemOpen
  \bibfield  {author} {\bibinfo {author} {\bibfnamefont {S.}~\bibnamefont
  {Teufel}}\ and\ \bibinfo {author} {\bibfnamefont {R.}~\bibnamefont
  {Tumulka}},\ }\href {\doibase 10.1007/s00220-005-1302-0} {\bibfield
  {journal} {\bibinfo  {journal} {Commun. Math. Phys.}\ }\textbf {\bibinfo
  {volume} {258}},\ \bibinfo {pages} {349} (\bibinfo {year}
  {2005})}\BibitemShut {NoStop}%
\bibitem [{\citenamefont {Sakurai}\ and\ \citenamefont
  {Commins}(2010)}]{sakurai2010modern}%
  \BibitemOpen
  \bibfield  {author} {\bibinfo {author} {\bibfnamefont {J.~J.}\ \bibnamefont
  {Sakurai}}\ and\ \bibinfo {author} {\bibfnamefont {E.~D.}\ \bibnamefont
  {Commins}},\ }\href@noop {} {\enquote {\bibinfo {title} {Modern quantum
  mechanics},}\ } (\bibinfo {year} {2010})\BibitemShut {NoStop}%
\bibitem [{\citenamefont {Holland}(2003)}]{Holland2003}%
  \BibitemOpen
  \bibfield  {author} {\bibinfo {author} {\bibfnamefont {P.~R.}\ \bibnamefont
  {Holland}},\ }\href {\doibase 10.1002/andp.200310022} {\bibfield  {journal}
  {\bibinfo  {journal} {Ann. Phys. (Leipzig)}\ }\textbf {\bibinfo {volume}
  {12}},\ \bibinfo {pages} {446} (\bibinfo {year} {2003})}\BibitemShut
  {NoStop}%
\bibitem [{\citenamefont {Holland}(1999)}]{Holland}%
  \BibitemOpen
  \bibfield  {author} {\bibinfo {author} {\bibfnamefont {P.~R.}\ \bibnamefont
  {Holland}},\ }\href {\doibase 10.1103/PhysRevA.60.4326} {\bibfield  {journal}
  {\bibinfo  {journal} {Phy. Rev. A}\ }\textbf {\bibinfo {volume} {60}},\
  \bibinfo {pages} {4326} (\bibinfo {year} {1999})}\BibitemShut {NoStop}%
\bibitem [{\citenamefont {D{\"{u}}rr}\ \emph {et~al.}(1992)\citenamefont
  {D{\"{u}}rr}, \citenamefont {Goldstein},\ and\ \citenamefont
  {Zangh{\`{I}}}}]{DGZBorn}%
  \BibitemOpen
  \bibfield  {author} {\bibinfo {author} {\bibfnamefont {D.}~\bibnamefont
  {D{\"{u}}rr}}, \bibinfo {author} {\bibfnamefont {S.}~\bibnamefont
  {Goldstein}}, \ and\ \bibinfo {author} {\bibfnamefont {N.}~\bibnamefont
  {Zangh{\`{I}}}},\ }\href {\doibase 10.1007/BF01049004} {\bibfield  {journal}
  {\bibinfo  {journal} {J. Stat. Phys.}\ }\textbf {\bibinfo {volume} {67}},\
  \bibinfo {pages} {843–} (\bibinfo {year} {1992})}\BibitemShut {NoStop}%
\bibitem [{\citenamefont {Corless}\ \emph {et~al.}(1996)\citenamefont
  {Corless}, \citenamefont {Gonnet}, \citenamefont {Hare}, \citenamefont
  {Jeffrey},\ and\ \citenamefont {Knuth}}]{Knuth}%
  \BibitemOpen
  \bibfield  {author} {\bibinfo {author} {\bibfnamefont {R.~M.}\ \bibnamefont
  {Corless}}, \bibinfo {author} {\bibfnamefont {G.~H.}\ \bibnamefont {Gonnet}},
  \bibinfo {author} {\bibfnamefont {D.~E.~G.}\ \bibnamefont {Hare}}, \bibinfo
  {author} {\bibfnamefont {D.~J.}\ \bibnamefont {Jeffrey}}, \ and\ \bibinfo
  {author} {\bibfnamefont {D.~E.}\ \bibnamefont {Knuth}},\ }\href {\doibase
  10.1007/BF02124750} {\bibfield  {journal} {\bibinfo  {journal} {Adv. Comput.
  Math.}\ }\textbf {\bibinfo {volume} {5}},\ \bibinfo {pages} {329} (\bibinfo
  {year} {1996})}\BibitemShut {NoStop}%
\bibitem [{\citenamefont {Moshinsky}(1952)}]{Moshinsky}%
  \BibitemOpen
  \bibfield  {author} {\bibinfo {author} {\bibfnamefont {M.}~\bibnamefont
  {Moshinsky}},\ }\href {\doibase 10.1103/PhysRev.88.625} {\bibfield  {journal}
  {\bibinfo  {journal} {Phys. Rev.}\ }\textbf {\bibinfo {volume} {88}},\
  \bibinfo {pages} {625} (\bibinfo {year} {1952})}\BibitemShut {NoStop}%
\bibitem [{\citenamefont {Goodman}(1981)}]{propagator}%
  \BibitemOpen
  \bibfield  {author} {\bibinfo {author} {\bibfnamefont {M.}~\bibnamefont
  {Goodman}},\ }\href {\doibase 10.1119/1.12720} {\bibfield  {journal}
  {\bibinfo  {journal} {American Journal of Physics}\ }\textbf {\bibinfo
  {volume} {49}},\ \bibinfo {pages} {843} (\bibinfo {year} {1981})}\BibitemShut
  {NoStop}%
\end{thebibliography}%

\end{document}